# A thermodynamic band gap model for photoinduced phase segregation in mixed-halide perovskites

Anthony Ruth[1,*], Halyna Okrepka[2], Prashant Kamat[2,3], Masaru Kuno[1,2,*]

1. University of Notre Dame, Department of Physics and Astronomy, Notre Dame, IN 46556, USA
2. University of Notre Dame, Department of Chemistry and Biochemistry and Department, Notre Dame, IN 46556, USA
3. Notre Dame Radiation Laboratory, Notre Dame, IN 46556, USA

aruth2@nd.edu
mkuno@nd.edu

**ORCID**
Anthony Ruth: 0000-0002-2670-5709
Halyna Okrepka: 0000-0002-3165-8521
Prashant Kamat: 0000-0002-2465-6819
Masaru Kuno: 0000-0003-4210-8514

**Abstract**

Provided is a comprehensive description of a band gap thermodynamic model, which predicts and explains key features of photosegregation in lead-based, mixed-halide perovskites. The model provides a prescription for illustrating halide migration driven by photocarrier energies. Where possible, model predictions are compared to experimental results. Free energy derivations are provided for three assumptions: (1) halide mixing in the dark, (2) a fixed number of photogenerated carriers funneling to and localizing in low band gap inclusions of the alloy, and (3) the statistical occupancy of said inclusions from a bath of thermalized photocarriers in the parent material. Model predictions include: excitation intensity ($I_{\mathrm{exc}}$)-dependent terminal halide stoichiometries ($x_{\mathrm{terminal}}$), excitation intensity thresholds ($I_{\mathrm{exc,threshold}}$) below which photosegregation is suppressed, reduced segregation in nanocrystals as compared to thin films, the possibility to kinetically manipulate photosegregation rates via control of underlying mediators, asymmetries in forward and reverse photosegregation rate constants/activation energies, and a preference for high band gap products to recombine with the parent phase. What emerges is a cohesive framework for understanding ubiquitous photosegregation in mixed-halide perovskites and a rational basis by which to manage the phenomenon.

## Introduction

Lead-based, mixed halide perovskites such as methylammonium lead iodide/bromide [$MAPb(I_{1-x}Br_x)_3$] are exciting materials that possess various light absorbing and light emitting applications. Of note is their use in tandem perovskite/silicon solar cells where certified efficiencies now exceed 29%.[1,2] The use of mixed-halide perovskites in applications is primarily motivated by their facile band gap ($E_g$) tuning, enabled through halide composition.[3] As an illustration, $MAPb(I_{1-x}Br_x)_3$ $E_g$-values can be tuned between 1.58 to 2.28 eV by varying $x$ from 0 to 1. This flexible electronic tuning remains true of other lead-based, mixed-halide perovskites given near edge transitions governed by valence bands composed of halide p and Pb 6s-orbitals.[4]

Unfortunately, mixed-halide perovskites possess an intrinsic instability beyond their well-documented environmental sensitivities.[5-9] This stems from illumination-induced halide photosegregation. First observed in $MAPb(I_{1-x}Br_x)_3$ by Hoke *et al.*[10], the phenomenon is seen as emission redshifts during illumination. Observed spectral shifting is attributed to $I^-$ and $Br^-$ photosegregation into I-rich and Br-rich inclusions. The former serves as energetically favorable regions for photogenerated carrier accumulation and subsequent radiative recombination. Structural confirmation comes from splitting of characteristic X-ray reflections into those associated with larger/smaller lattice constant I-rich and Br-rich domains. Most impressive about the phenomenon is that the original mixed-halide perovskite is restored through subsequent, entropically-driven, dark remixing.

Illumination-induced halide demixing and photosegregation has since been observed in other mixed-halide, hybrid and all-inorganic perovskites. Systems where the phenomenon has been seen include formamidinium lead iodide/bromide [$FAPb(I_{1-x}Br_x)_3$],[11] methylammonium lead chloride/bromide [$MAPb(Cl_{1-x}Br_x)_3$],[12] cesium lead iodide/bromide [$CsPb(I_{1-x}Br_x)_3$],[13] and even in mixed-cation/mixed-anion systems such as $(MA,FA,Cs)Pb(I_{1-x}Br_x)_3$.[14-16]

To rationalize the phenomenon and possibly mitigate its effects, various models and theoretical constructs have been developed. They include thermodynamic[17,19,20,21], chemical[22,23,24], polaron[25,26,27], and trap-related[28,29,30,31] models. In short, thermodynamic models include miscibility gap[17] and band gap-based variants.[19,20] The former attributes photosegregation to poor anion mixing that results in metastable alloys prone to spinodal decomposition when perturbed.[17] The latter is described below.

Chemical models suggest facile iodide oxidation due to photogenerated holes. This results in iodine migration via oxidized chemical species such as $I_2$ and $I_3^-$ as well as by interstitials (*e.g.*, $I_i^+$ and $I_i$). Spatially heterogeneous iodine depletion/subsequent lattice accumulation produces local I-rich and Br-rich domains. Polaron models, in contrast, posit that strong electronic-ionic interactions localize photogenerated charges. Resulting polarons expand lattices and lead to local anion displacements due to anion-specific preferential bond lengths. Trap-related models alternatively assume electric fields from localized charges that drive point defect migration and associated anion migration. References 32-34 review photoinduced halide segregation in lead-based, hybrid/all-inorganic perovskites and describe how existing models account for known experimental observations.

In what follows, we summarize a band gap-based thermodynamic model of halide photosegregation that rationalizes nearly all experimental aspects of photoinduced halide demixing.[19,20,33] At its most basic level, the model suggests that local band gap differences between parent, mixed-halide alloys and stochastically-generated or light-induced iodine-rich regions provide a thermodynamic driving force to induce halide photosegregation. Band gap sensitivities appear in excess photocarrier energies that enable vacancy-mediated[20] anion hopping within perovskite lattices. This, in turn, enables photoexcited, mixed-halide perovskites to reduce their free energies by creating segregated iodine- and bromine-rich domains. In the absence of above-gap excitation, the original lattice restores itself through entropically-driven, anion remixing.

## Model

What follows is a cogent and relatively complete description of the band gap thermodynamic model. The first step involves formulating a photosegregation reaction that produces an iodine-rich and bromine-rich phase in a parent, mixed-phase alloy. The chosen reaction is

$$\mathrm{APb}\left(\mathrm{I}_{1-\mathrm{x}_{\mathrm{init}}}\mathrm{Br}_{\mathrm{x}_{\mathrm{init}}}\right)_3 \leftrightarrow \frac{1}{2}\mathrm{APb}(\mathrm{I}_{1-\mathrm{x}}\mathrm{Br}_{\mathrm{x}})_3 + \frac{1}{2}\mathrm{APb}\left(\mathrm{I}_{1-2\mathrm{x}_{\mathrm{init}}+x}\mathrm{Br}_{2\mathrm{x}_{\mathrm{init}}-\mathrm{x}}\right)_3 \tag{1}$$

where A=methylammonium ($MA^+$) or formamidinium ($FA^+$) or $Cs^+$, $x_{\mathrm{init}}$ is the alloy's initial stoichiometry, and the first (second) product represents an I-rich (Br-rich) product phase. To simplify subsequent thermodynamic analysis, **Equation 1** is alternatively expressed as

$$(1 - x_{\mathrm{init}})\mathrm{APbI}_3 + x_{\mathrm{init}}\mathrm{APbBr}_3 \leftrightarrow \frac{1}{2}\mathrm{APb}(\mathrm{I}_{1-x}\mathrm{Br}_{x})_3 + \frac{1}{2}\mathrm{APb}\left(\mathrm{I}_{1-2x_{\mathrm{init}}+x}\mathrm{Br}_{2x_{\mathrm{init}}-x}\right)_3 \tag{2}$$

where the mixed-halide alloy, $\mathrm{APb}\left(\mathrm{I}_{1-\mathrm{x}_{\mathrm{init}}}\mathrm{Br}_{\mathrm{x}_{\mathrm{init}}}\right)_3$, has been written in terms of its starting pure halide precursors. Of note in either **Equations 1** or **2** are equal quantities of I-enriched and Br-enriched products.[16,19,35,36] As will be described next, other chemical descriptions of photosegregation exist.

Although product symmetries in **Equations 1** and **2** make them particularly convenient to model, there are alternative chemical reactions, which lead to (asymmetric) product stoichiometries. Their validity is supported by recent, compositionally-weighted X-ray diffraction studies that reveal wide distributions of product stoichiometries underlying mixed-halide photosegregation.[37] Supporting Information (**SI**) **Note 1** summarizes alternative mixing and segregation reactions. **SI Note 2** provides analytical expressions for their mixing entropies and enthalpies. The choice of which mixing/segregation reaction to use is motivated by how well they simplify free energy expressions, whether their free energies are smaller than those of other expressions, and whether they introduce/use key photosegregation parameters such as initial halide and variable product stoichiometries.

In what follows, **Equation 2** is assumed to simplify analytical modeling of (dark) enthalpic and entropic mixing energies. A corresponding (halide) mixing free energy ($\Delta F_{\mathrm{mix}}$) is

$$\Delta F_{\mathrm{mix}} = \Delta H_{\mathrm{mix}} - T\Delta S_{\mathrm{mix}} \tag{3}$$

where $\Delta H_{\mathrm{mix}}$ ($\Delta S_{\mathrm{mix}}$) is the enthalpy (entropy) of mixing. When illuminated (*), the overall mixing free energy becomes

$$\Delta F^* = \Delta F_{\mathrm{mix}} + \Delta F_{\mathrm{light}} \tag{4}$$

where $\Delta F_{\mathrm{light}}$ is a term that stems from irradiation of the mixed-halide alloy. $\Delta F_{\mathrm{light}}$ accounts for energetic changes that occur due to carrier funneling and accumulation in either stochastically- or photo-generated I-rich inclusions of the parent alloy. That carriers funnel to and accumulate within I-rich domains is rationalized by favorable parent alloy/I-rich inclusion band offsets[19,38] as well as by long carrier diffusion lengths between 100-1000 nm in perovskites.[19,36,39,40,41,42]

Because carrier thermalization in I-rich domains is responsible for excess energies that promote anion migration, the band gap thermodynamic model considers local band gap gradients between parent, mixed-halide alloys and their I-rich inclusions. In practice, a Vegard's law expression,

$$E_{\mathrm{PL}}(x) = E_{\mathrm{PL},x=0}(1 - x) + E_{\mathrm{PL},x=1}x - bx(1 - x), \tag{5}$$

describes band edge photoluminescence (PL) energies ($E_{\mathrm{PL}}$) [3] with $E_{\mathrm{PL},x=0}$ and $E_{\mathrm{PL},x=1}$ bulk,

temperature-dependent PL energies of limiting $APbI_3$ and $APbBr_3$ stoichiometries. $E_{PL}$ is used as a proxy for the system's band gap, ($E_g$), given the preponderance of emission-based photosegregation measurements. $b$ is a bowing parameter that accounts for changes to mixed-halide perovskite band energies due to compositional disorder. When employed within the context of analogous Vegard's law expressions of perovskite lattice constants,[32] $b$ captures extant bonding anisotropies that stem from lattice mismatches between limiting compositions.

**Figure 1a** summarizes the Vegard's law $E_{PL}$ behavior of $MAPb(I_{1-x}Br_x)_3$, $FAPb(I_{1-x}Br_x)_3$, and $CsPb(I_{1-x}Br_x)_3$. Highlighted are band offsets that favor carrier accumulation in I-rich inclusions of the parent alloy.[19,38] Corresponding composition-dependent band gaps are illustrated in **Figure 1b** with **Table 1** summarizing literature Vegard's law $E_{PL}$ parameters for these and other lead-based, mixed-halide perovskites.

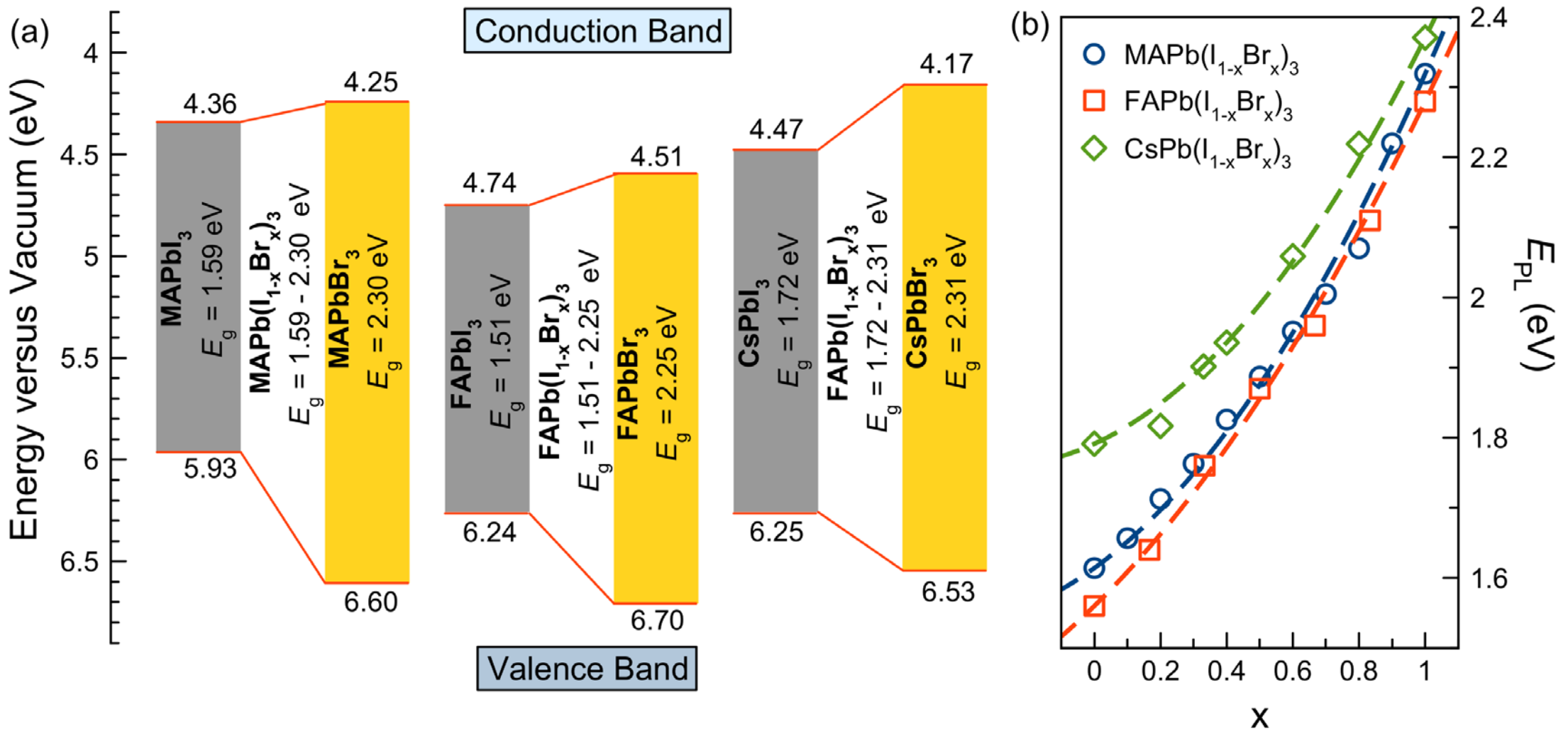

**Figure 1.** (a) $APbI_3$ and $APbBr_3$ band offsets relative to vacuum (A=$MA^+$, $FA^+$ and $Cs^+$). Data extracted from Reference 38. (b) Composition-dependent $E_{PL}$ for all three materials. Dashed lines are fits to Vegard's law expressions from References 10, 43, and 44.

**Table 1**. Vegard's law $E_{PL}$ parameters for various lead-based, mixed-halide perovskites.

| System | $E_{PL,x=0}$ (eV) | $E_{PL,x=1}$ (eV) | $b$ (eV) | Reference |
|---|---|---|---|---|
| $MAPb(I_{1-x}Br_x)_3$ | 1.61 | 2.30 | 1.042 | 44 |
| | 1.58 | 2.30 | 0.324 | 16 |
| | 1.59 | 2.33 | 0.095 | 45 |
| | 1.61 | 2.32 | 0.364 | 10 |
| $FAPb(I_{1-x}Br_x)_3$ | 1.56 | 2.28 | 0.259 | 44 |
| | 1.52 | 2.27 | -0.029 | 46 |
| $CsPb(I_{1-x}Br_x)_3$ | 1.79 | 2.37 | 0.364 | 43 |
| $MACsPb(I_{1-x}Br_x)_3$ | 1.67 | 2.33 | 0.310 | 16 |
| $FAMAPb(I_{1-x}Br_x)_3$ | 1.59 | 2.28 | 0.420 | 44 |
| $FACsPb(I_{1-x}Br_x)_3$ | 1.54 | 2.25 | 0.363 | 47 |
| | 1.58 | 2.28 | 0.354 | 14 |
| $FAMACsPb(I_{1-x}Br_x)_3$ | 1.58 | 2.27 | 0.092 | 48 |

Evaluating **Equation 4** requires considering $\Delta F_{mix}$ (*i.e.*, $\Delta H_{mix}$, $\Delta S_{mix}$) and $\Delta F^*$ on equal footings. $\Delta H_{mix}$ and $\Delta S_{mix}$ have been described previously and are common to both band gap and polaron models.[16] $\Delta H_{mix}$, in particular, encompasses inter-anion repulsive or attractive interactions that Chen *et al.*[21] suggest are of order 8-12 meV per halide atom at room temperature.

$\Delta H_{mix}$ can be modeled using an analytical expression of the form

$$\Delta H_{\text{mix}}(x) = U_{\text{I,Br}}x(1-x) \tag{6}$$

where $U_{I,Br}$ is a bimolecular I-Br interaction energy of order $U_{I,Br}$~136 meV per formula unit (FU) or ~45 meV per halide atom for $CsPb(I_{1-x}Br_x)_3$. For compositions such as $MA(I_{1-x}Br_x)_3$ and $MA_{0.88}Cs_{0.12}Pb(I_{1-x}Br_x)_3$, $U_{I,Br}$=39 and $U_{I,Br}$=44 meV per halide atom respectively. In the case of $FA(I_{1-x}Br_x)_3$ and $FA_{0.88}Cs_{0.12}Pb(I_{1-x}Br_x)_3$, analytical approximations for $\Delta H_{mix}$ require additional cubic terms, *i.e.*,

$$\Delta H_{\text{mix}}(x) = \frac{U'_{\text{I,Br}}}{2}\left[x(1-x)^2 - x^2(1-x)\right] \tag{7}$$

with $U'_{I,Br}$=103 and $U'_{I,Br}$=87 meV per halide atom respectively. Both $U_{I,Br}$ and $U'_{I,Br}$ are weakly temperature dependent, allowing their $T$-dependencies to be neglected. From **Equation 6**, a total mixing enthalpy for **Equation 2** is

$$\Delta H_{\text{mix}}(x) = -U_{\text{I,Br}}(x - x_{\text{init}})^2 + U_{\text{I,Br}}x_{\text{init}}(1 - x_{\text{init}}). \tag{8}$$

Corresponding entropic contributions are expressed using Shannon entropy[49] where for a given reactant or product

$$S = -k \times [A\ln(A) + B\ln(B)]. \tag{9}$$

In **Equation 9**, $k$ is the Boltzman constant, $A$ is the site occupancy of species $A$ (*e.g.*, $I^-$), and $B$ is the site occupancy of species $B$ (*e.g.*, $Br^-$). **Equation 2**'s entropic change is therefore

$$T\Delta S_{\text{mix}} = -\frac{3}{2}kT[(2x_{\text{init}} - x)\ln(2x_{\text{init}} - x) + x\ln(x) + (1 + x - 2x_{\text{init}})\ln(1 + x - 2x_{\text{init}}) + (1-x)\ln(1-x)]. \tag{10}$$

**Equation 10** can be simplified by taking a second order Taylor series expansion about $x$=$x_{init}$ [*i.e.*, $T\Delta S_{\text{mix}} \approx \left(T\Delta S_{\text{mix}}|_{x=x_{\text{init}}} + (x - x_{\text{init}})\frac{dT\Delta S_{\text{mix}}}{dx}|_{x=x_{\text{init}}} + (x - x_{\text{init}})^2\frac{1}{2}\frac{d^2T\Delta S_{\text{mix}}}{dx^2}|_{x=x_{\text{init}}}\right)$]. The result is expressed on a per halide basis by dividing by three to obtain

$$-T\Delta S_{\text{mix}} \approx \frac{kT}{2}\frac{1}{x_{\text{init}}(1-x_{\text{init}})}\{(x - x_{\text{init}})^2 - x_{\text{init}}^2\}. \tag{11}$$

At this point, what distinguishes the various models and exclusive to the band gap thermodynamic model is $\Delta F_{light}$. This term accounts for photogenerated carrier contributions to $\Delta F^*$ via carrier funneling and localization in I-rich domains of the lattice. Of particular importance is the $E_g$ difference between the parent alloy and its I-rich inclusions, as captured by gradients in **Equation 5**. Equally important is how one defines local $E_g$-values of the material and, specifically, the number of anions that must migrate to change $E_g$ substantially. In what follows, both are addressed by defining a finite volume of the parent alloy, $V_{\text{gap}}$, that constitutes a domain having $j = \frac{V_{\text{gap}}}{V_{\text{u.c.}}}$ formula units. $V_{\text{u.c.}}$ is the volume of an individual unit cell of the lattice.

$V_{\text{gap}}$ has previously been estimated from increases in PL linewidths of mixed-halide perovskite alloys relative to those of their pure iodide or pure bromide counterparts.[36] Conceptually, electronic disorder in mixed-halide alloys is linked to stochastic variations in halide stoichiometry within $V_{\text{gap}}$ elements of the lattice. What results is the following statistical expression for $V_{\text{gap}}$,

$$V_{\text{gap}} \geq \frac{V_{\text{u.c.}}}{3}\left[\left(\frac{dE_{\text{g}}}{dx}\right)\frac{\sqrt{x(1-x)}}{\delta E_{\text{g}}}\right]^2 \tag{12}$$

that is derived in **SI Note 3**. Introducing the following parameters for $MAPb(I_{1-x}Br_x)_3$ [$CsPb(I_{1-x}Br_x)_3$], $V_{\text{u.c.}}$=0.257 [0.227] nm$^3$, $\frac{dE_{\text{g}}}{dx}$=1.11 [0.6775] eV, $\delta E_{\text{g}}$=80 [60] meV, and $x$=0.5, yields $V_{\text{gap}}$~4.1 [2.4] nm$^3$. Corresponding radii are $r_{\text{gap}}$~ 1.00 and 0.83 nm, respectively.

An alternate approach for estimating $V_{\text{gap}}$ utilizes the exciton Bohr radius ($a_{\text{B}}$) of perovskites. **Table 2** summarizes literature-reported $a_{\text{B}}$-values for various Pb-based hybrid and all-inorganic perovskites. Underlying this approach is that any domain, which determines the local band gap of a perovskite, must remain charge neutral. Consequently, even if band offsets induce a photogenerated charge to localize in an I-rich domain, an opposite charge will be attracted to its vicinity to effectively neutralize it. This necessitates a lattice volume, defined by $a_{\text{B}}$, where Coulombic interactions in the lattice remain meaningful. Reported perovskite $a_{\text{B}}$-values between 3.5-6.4 nm therefore suggest $V_{\text{gap}}$-values of order 270 nm$^3$. Empirically, $V_{\text{gap}}$-values, estimated using electronic disorder, better match experimental photosegregation behavior.

**Table 2**. Literature $a_{\text{B}}$-values for Pb-based hybrid and inorganic perovskites.

| System | $a_B$ (nm) | Reference |
|---|---|---|
| $MAPbBr_3$ | 3.8 | 50 |
| $MAPbI_3$ | 4.6 | 51 |
| $FAPbBr_3$ | 3.5 | 52 |
| $FAPbI_3$ | 6.4 | 52 |
| $CsPbBr_3$ | 3.5 | 53 |
| $CsPbI_3$ | 6.0 | 53 |

Having established $V_{\text{gap}}$, $I^-$ enrichment of an I-rich inclusion means a corresponding decrease of its $Br^-$ content by $\delta x = \frac{1}{3j}$.[20,36] The factor of three accounts for three unique anions per unit cell. In tandem, the parent phase, which consists of $m$ formula units ($m \gg j$), experiences a bromine fraction change of $\delta x' = \frac{1}{3m}$. Qualitatively, I-rich domains exchange $Br^-$ with $I^-$ from their parent, mixed-halide phase via

$$j\ \text{APb}(\text{I}_{1-x}\text{Br}_x)_3 + m\ \text{APb}\left(\text{I}_{1-x_{\text{init}}}\text{Br}_{x_{\text{init}}}\right)_3 \leftrightarrow j\ \text{APb}(\text{I}_{1-x+\delta x}\text{Br}_{x-\delta x})_3 + m\ \text{APb}\left(\text{I}_{1-x_{\text{init}-\delta x'}}\text{Br}_{x_{\text{init}+\delta x'}}\right)_3. \tag{13}$$

**Equation 13** differs from **Equations 1** and **2** by how photosegregation is modeled. Whereas the latter focuses on the demixing of a parent alloy, which results in the (symmetric) production of I-rich and Br-rich inclusions, **Equation 13** models the microscopic exchange of $I^-$ and $Br^-$ between an I-rich inclusion and its parent alloy. When using **Equation 1**, a terminal photosegregation stoichiometry (discussed below in more detail) is associated with a $\Delta F^*$ minimum (*i.e.*, $\frac{d\Delta F^*}{dx} = 0$). For **Equation 13**, terminal stoichiometries are associated with the condition where stepwise enrichment no longer becomes favorable. This occurs when its free energy crosses zero. That **Equations 1 (2)** and **Equation 13** lead to near identical photosegregation behavior can be verified by identical $\frac{d\Delta F_{\text{mix}}}{dx} = 0$ for **Equation 1** and $\Delta F_{\text{mix}}=0$ for **Equation 13**.

**Equation 13** is therefore used to model $\Delta F_{\text{light}}$ by simultaneously considering the energetic change incurred by $I^-$ enrichment of individual domains. The number of carriers, $n$, involved in this process is

$$n \leq \frac{I_{\text{exc}}\alpha}{h\nu}\tau V_D \tag{14}$$

where $\alpha$ is a perovskite absorption coefficient, $h\nu$ is the incident photon energy, $\tau$ is the excited state lifetime, and $V_D = \frac{4}{3}\pi l_{\text{e/h}}^3$ is a diffusion volume with $l_{\text{e/h}}$, a corresponding electron or hole diffusion length. **Equation 13** assumes that only one I-rich domain exists within a carrier's diffusion volume. In practice, multiple domains likely exist within $V_D$ given large $l_{\text{e/h}}$-values. This illustrates a current simplification of the model.

Energetic changes incurred by $I^-/Br^-$ exchange are parameterized by $\Delta E_{\text{g,domain}}$[16,20,36] where

$$\Delta E_{\text{g,domain}} = \frac{dE_{\text{g}}}{dx}\delta x \tag{15}$$

with

$$\frac{dE_{\text{g}}}{dx} = \left(E_{\text{PL},x=1} - E_{\text{PL},x=0} - b\right) + 2bx \tag{16}$$

obtained using **Equation 5**. In turn, **Equations 14-16** yield the following macroscopic free energy change from photogenerated carrier-induced anionic rearrangements

$$\Delta F_{\text{light}}(x) = n\Delta E_{\text{g,domain}}(x - x_{\text{init}}). \tag{17}$$

**Equation 17** is then combined with $\Delta H_{\text{mix}}$ (**Equation 8**) and $\Delta S_{\text{mix}}$ (**Equation 11**) to yield the following explicit form for $\Delta F^*$.

$$\Delta F^*(x) = (\Delta H_{\text{mix}} - T\Delta S_{\text{mix}}) + \Delta F_{\text{light}}(x)$$
$$= \left\{-U_{\text{I,Br}} + \frac{kT}{2}\left[\frac{1}{x_{\text{init}}(1-x_{\text{init}})}\right]\right\}(x - x_{\text{init}})^2 + U_{\text{I,Br}}x_{\text{init}}(1 - x_{\text{init}}) - \frac{kT}{2}\left[\frac{1}{x_{\text{init}}(1-x_{\text{init}})}\right]x_{\text{init}}^2 + n\Delta E_{\text{g,domain}}(x - x_{\text{init}}). \tag{18}$$

The importance of $\Delta F_{\text{light}}$ in establishing the overall behavior of **Equation 18** is higlighted by its magnitude. Assuming the following parameters for $MAPb(I_{1-x}Br_x)_3$ [$CsPb(I_{1-x}Br_x)_3$], $x_{\text{init}}=0.5$, a nominal composition change of ($x_{\text{init}}$-$x$)=0.5, $U_{\text{I,Br}}$=39 meV [45 meV], $V_{\text{gap}}$=4.1 nm$^3$ [2.4 nm$^3$], $V_{\text{u.c.}}$=0.252 nm$^3$ [0.250 nm$^3$], $I_{\text{exc}}$=100 mW cm$^{-2}$, $\alpha$=10$^5$ cm$^{-1}$ [54], $h\nu$=3 eV, $\tau$=100 ns[54], and $V_D$=4.189 μm$^3$ (corresponding carrier diffusion length, $l_{\text{e/h}}$=1000 nm), **Equation 17** yields a (liberal maximum) photoexcitation contribution to $\Delta F^*$ of order 100 eV. Other combinations of $\tau$ and $l_{e/h}$, along with the above parameters, readily yield $\Delta F^*$-values of order 1 eV. This should be contrasted to associated mixing enthalpies (**Equation 8**) and entropies (**Equation 11**) of order +10 and -20 meV, respectively. Evident is that electronic contributions dominate $\Delta F^*$ in a band gap thermodynamic model and tip mixing free energies towards demixing when $n$>0.

**Figure 2** now illustrates $\Delta F^*$ for the band gap thermodynamic model as a function of $n$ for $x_{init}$=0.5. Evident is that in the dark, the free energy minimum occurs at $x_{init}$. At low light intensities and above a threshold intensity (derived below), $\Delta F^*$'s minimum shifts towards lower bromine content. Above $n$=0.2, $\Delta F^*$ decreases monotonically with $x$ and effectively follows the Vegard's law composition dependent band gap of the mixed-halide perovskite. This results in a corresponding free energy minimum at $x \approx 0$. In short, over a relatively narrow window in $n$ of approximately 1 order of magnitude, the system smoothly transitions between being stable in the dark to completely favoring I-rich domain formation/growth. Analogous behavior is seen with other $x_{init}$.

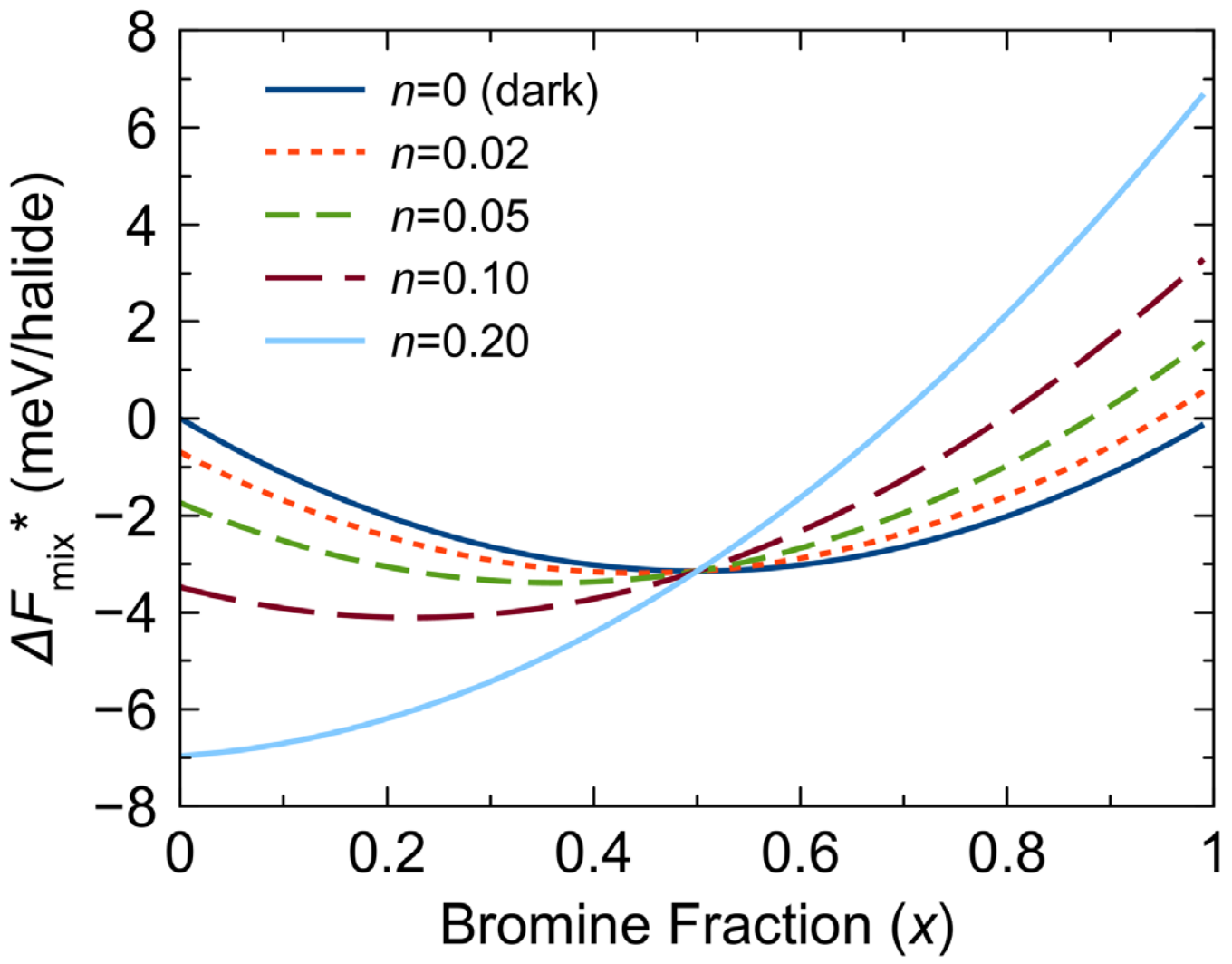

**Figure 2**. $\Delta F^*$ from a thermodynamic band gap model versus $x$ for $x_{init}$=0.5 and $n$=0 (dark), 0.02, 0.05, 0.10, and 0.2.

One of the more important results to emerge from **Equation 18** as well as evident in **Figure 2** is an explicit prediction of $I_{exc}$-dependent $x_{terminal}$ in the *low intensity* limit (*i.e.*, small $n$).[36] $x_{terminal}$ is found by evaluating critical points in $\Delta F^*$ for small departures about $x_{init}$. What results is[16,36]

$$x_{terminal} = x_{init}\left[1 - (1 - x_{init})\left(\frac{n\Delta E_{g,domain}}{kT - 2U_{I,Br}x_{init}(1 - x_{init})}\right)\right] \quad (19)$$

where $n$ (**Equation 14**) encompasses $x_{terminal}$'s $I_{exc}$ dependence. Of note is the dearth of work investigating this dependency. It is likely that prior analyses have assumed identical (final) photosegregated states.[19,29,30,35,55]

Recent work, conducted with $CsPb(I_{1-x}Br_x)_3$ nanocrystals and thin films, however, reveal apparent $x_{terminal}$ $I_{exc}$-dependencies.[36] This is highlighted in **Figure 3a**, which plots peak photosegregation wavelengths and energies observed during photosegregation. Peak wavelengths (energies) increase (decrease) until eventually saturating on a ~1 hour timescale. More importantly, terminal energies ($E_{terminal}$) appear $I_{exc}$-dependent and decrease with increasing $I_{exc}$. **Figure 3b** shows analogous behavior for a 25 nm edge length ($l_{nanocrystal}$) nanocrystal ensemble. To ensure that observed spectral changes do not arise from irreversible photochemistries linked to iodine loss[22,56,57], all data have been checked for reversibility.

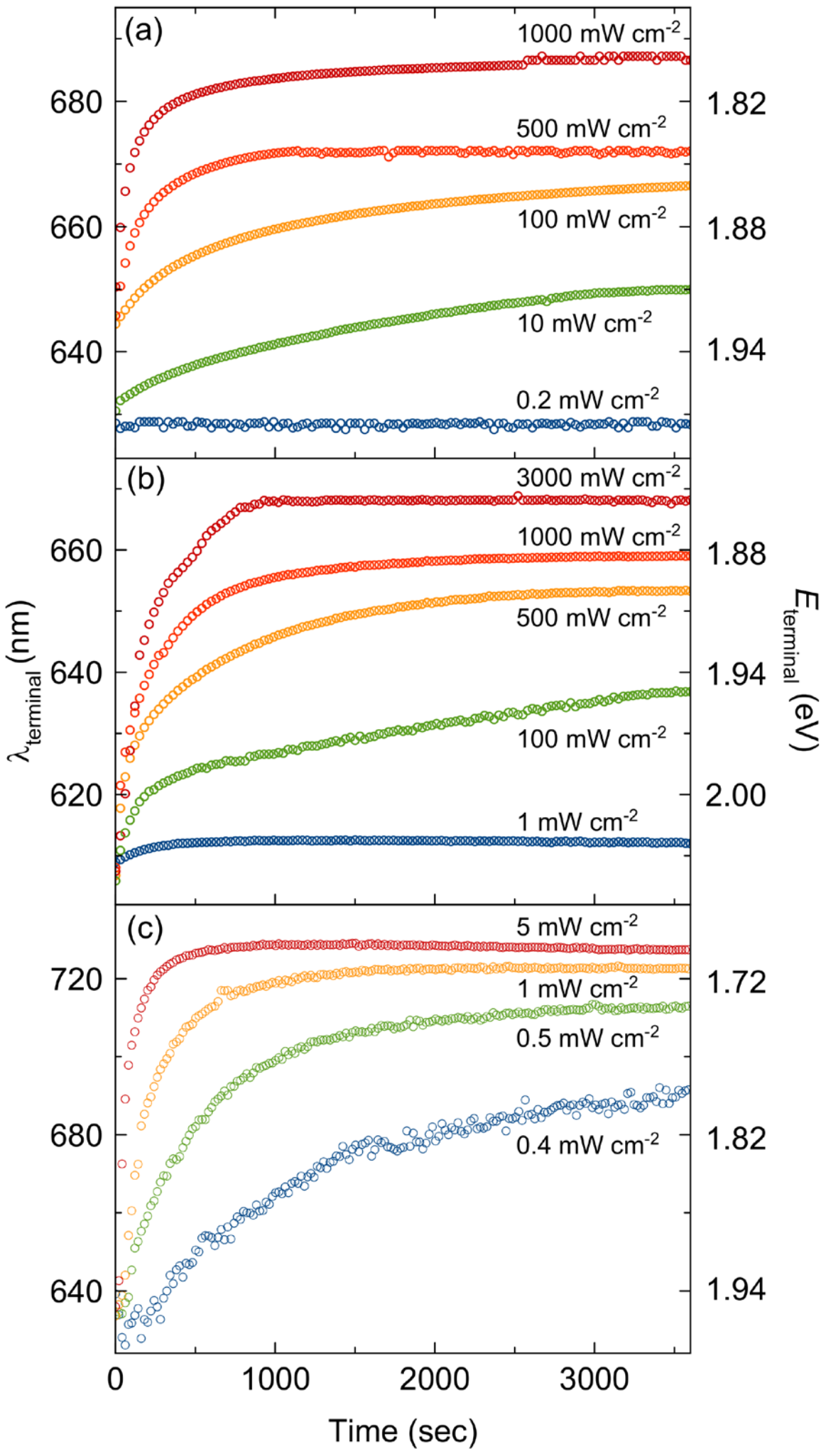

**Figure 3**. $I_{exc}$-dependent photosegregation emission wavelengths/energies for (a) a $CsPb(I_{0.5}Br_{0.5})_3$ thin film and (b) a $l_{nanocrystal}$=25 nm $CsPb(I_{0.5}Br_{0.5})_3$ nanocrystal ensemble. Data from Reference 36. (c) Analogous data for a $MAPb(I_{0.5}Br_{0.5})_3$ thin film.

**Figure 3c** shows data acquired on a $MAPb(I_{0.5}Br_{0.5})_3$ thin film where PL peak wavelengths have been plotted as a function of time. Of note are the very low excitation intensities used to demonstrate apparent $I_{exc}$-dependent saturation of $E_{terminal}$. This possibly explains why prior $MAPb(I_{1-x}Br_x)_3$ studies have overlooked this effect, given experimental excitation intensities often exceeding 10 mW $cm^{-2}$.[19,29,30]

All $E_{terminal}$ in **Figure 3** can be converted to corresponding $x_{terminal}$ using Vegard's law expressions of composition-dependent energies (**Table 1**). For the $CsPb(I_{1-x}Br_x)_3$ thin film (**Figure 3a**), this results in $x_{terminal}$-values that range from $x_{terminal}$=0.46 to 0.05 for $I_{exc}$=0.2-1000 mW $cm^{-2}$. For the $l_{nanocrystal}=l_{e/h}$=25 nm nanocrystal ensemble (**Figure 3b**), $x_{terminal}$-values range from $x_{terminal}$=0.56 to 0.21 for $I_{exc}$=0.5-3000 mW $cm^{-2}$. For the $MAPb(I_{1-x}Br_x)_3$ thin film (**Figure 3c**), $x_{terminal}$-values range from $x_{terminal}$=0.41 to 0.22 for $I_{exc}$=0.4-100 mW $cm^{-2}$.

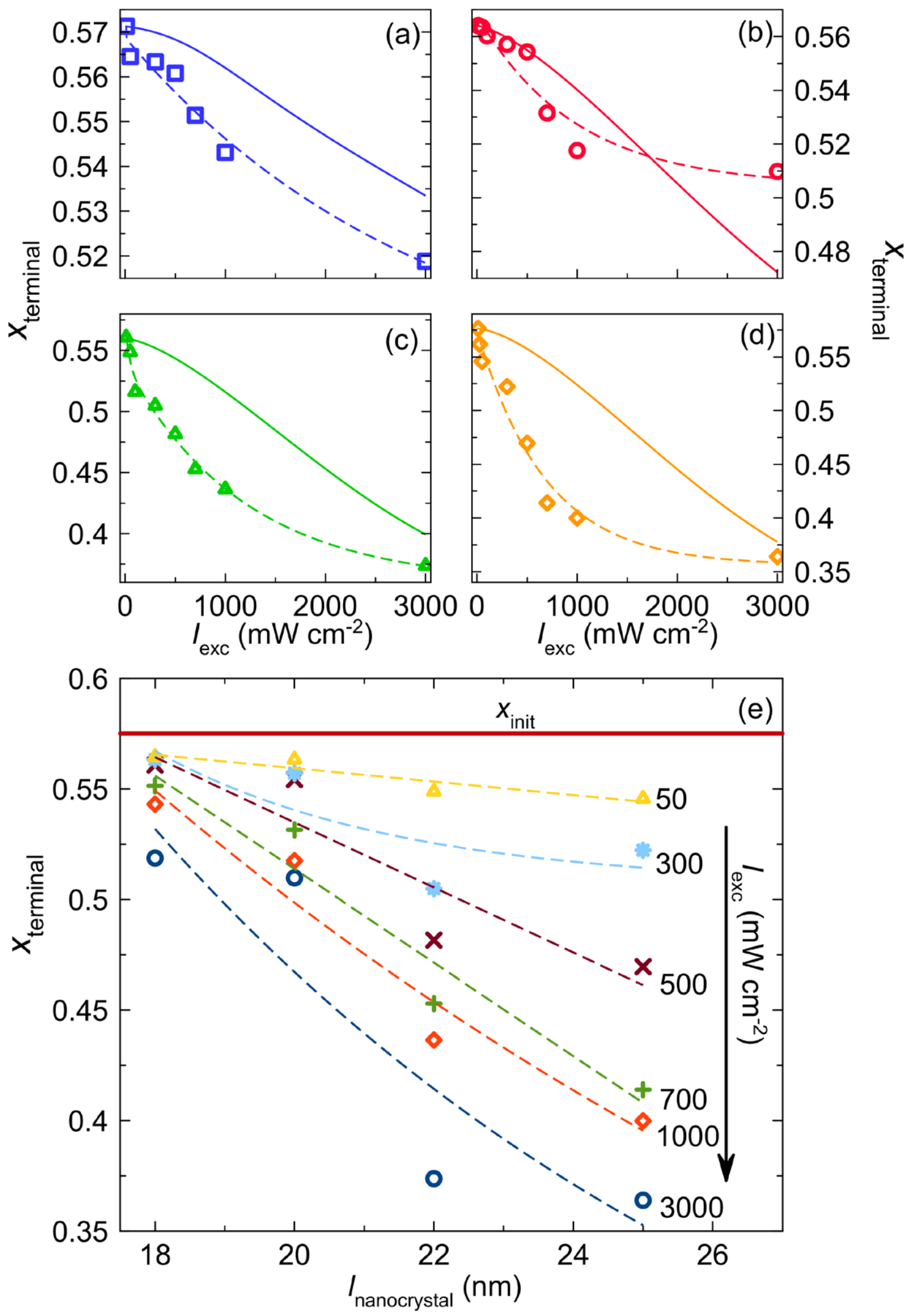

**Figure 4**. $CsPb(I_{0.5}Br_{0.5})_3$ nanocrystal ensemble $x_{terminal}$-values acquired at different $I_{exc}$ for $l_{nanocrystal}$= (a) 18 nm, (b) 20 nm, (c) 22 nm, and (d) 25 nm. Solid lines are theoretical predictions from **Equation 19**. Data from Reference 36. (e) $x_{terminal}$-values as functions of $l_{nanocrystal}$ for a given $I_{exc}$. Solid line, $x_{init}$. In all cases, dashed lines are guides to the eye.

**Figures 4 a-d** summarize extracted $x_{terminal}$-values for $CsPb(I_{1-x}Br_x)_3$ nanocrystals that highlight $x_{terminal}$'s $I_{exc}$ dependency. Superimposed atop the data are predictions from **Equation 19** (solid lines). A corresponding comparison has not been made for either $CsPb(I_{1-x}Br_x)_3$ or $MAPb(I_{1-x}Br_x)_3$ thin film data. This is because whereas $l_{e/h}$ in nanocrystals is defined by nanocrystal edge lengths (*i.e.*, $l_{e/h} \approx l_{nanocrystal}$), in bulk materials, $l_{e/h}$ refers to carrier diffusion lengths that are heavily influenced by grain sizes, defect densities and by the photonucleation of new I-rich domains within the same carrier diffusion volume, $V_D$.

**Figures 4 a-d** show that when $l_{e/h}$ (*i.e.*, $V_D$) is well defined **Equation 19** captures the decrease in $x_{terminal}$ with increasing $I_{exc}$. The model, however, fails to fully capture observed $I_{exc}$ dependencies. This is best highlighted by qualitative differences in experimental versus predicted $x_{terminal}$ curvatures with increasing $I_{exc}$. The former exhibits a concave up behavior while the latter bends concave down. The experiment therefore shows a greater $x_{terminal}$ sensitivity to $I_{exc}$ at low to moderate excitation intensities than predicted by theory.

Model comparisons to experiment are complicated by $x_{terminal}$ co-dependencies in **Equation 19**. In particular, $x_{terminal}$ depends on $\tau$, an effective carrier lifetime in photosegregated or photonucleated I-rich domains. Comparisons in **Figures 4 a-d** have therefore entailed approximating $\tau$ using PL lifetime measurements of photosegregated emission in nanocrystals.[36] $x_{terminal}$ additionally depends on nanocrystal size or bulk carrier diffusion length. This is highlighted in **Figure 4e** where $x_{terminal}$-values have been plotted as functions of $l_{nanocrystal}$ for a given $I_{exc}$. As expected, smaller $l_{nanocrystal}$-values suppress photosegregation.

Despite the overall lack of quantitative agreement between theory and experiment, the band gap thermodynamic model captures observed experimental trends in **Figure 4**. Improving the theory/experiment comparison likely requires generalizing **Equation 2** to include product distributions seen in recent compositionally-weighted X-ray diffraction measurements.[37] Additionally, photonucleation of multiple I-rich domains within given I-rich inclusions should be considered while explicitly considering the compositional sensitivity of $V_{gap}$.

Beyond this, **Equation 18** explicitly considers $\Delta F^*$ for the assumed reaction in **Equation 2**. Emission-based experiments, however, report the PL of minority species with the smallest band gaps. This (~1-2%) minority fraction is a subpopulation of an already small (~5-23%) fraction of the parent alloy undergoing photosegregation.[10,55] What results is an experimental bias towards reporting the photoresponse of the lowest $x$ domains. Experimental $x_{terminal}$-values will therefore be smaller than those predicted by the band gap thermodynamic model.

Although applying **Equation 19** to bulk materials is complicated by uncertainties in $l_{e/h}$, it is possible to check its consistency. **Figure 3a** has already shown $I_{exc}$-dependent $E_{terminal}$-values for $CsPb(I_{1-x}Br_x)_3$ thin films. **Figure 5** now plots corresponding $I_{exc}$-dependent $x_{terminal}$-values from a Vegard's law conversion of $E_{terminal}$ (**Table 1**). Plotted alongside the experimental data are estimated $l_{e/h}$-values from **Equation 19**, required to arrive at the same $x_{terminal}$. Model estimates incorporate experimentally-measured (bulk) $\tau$-values at each $I_{exc}$.[36] Extracted $l_{e/h}$ range from $l_{e/h}$=27-129 nm and agree with reported perovskite carrier diffusion lengths.[19,36,39,40,41,42] Observed decreases of model-extracted $l_{e/h}$-values with increasing $I_{exc}$, in turn, point to the nucleation of additional $I$-rich domains and suggest that higher excitation intensities favor increasing the number of I-rich domains as opposed to expanding the size of existing inclusions.

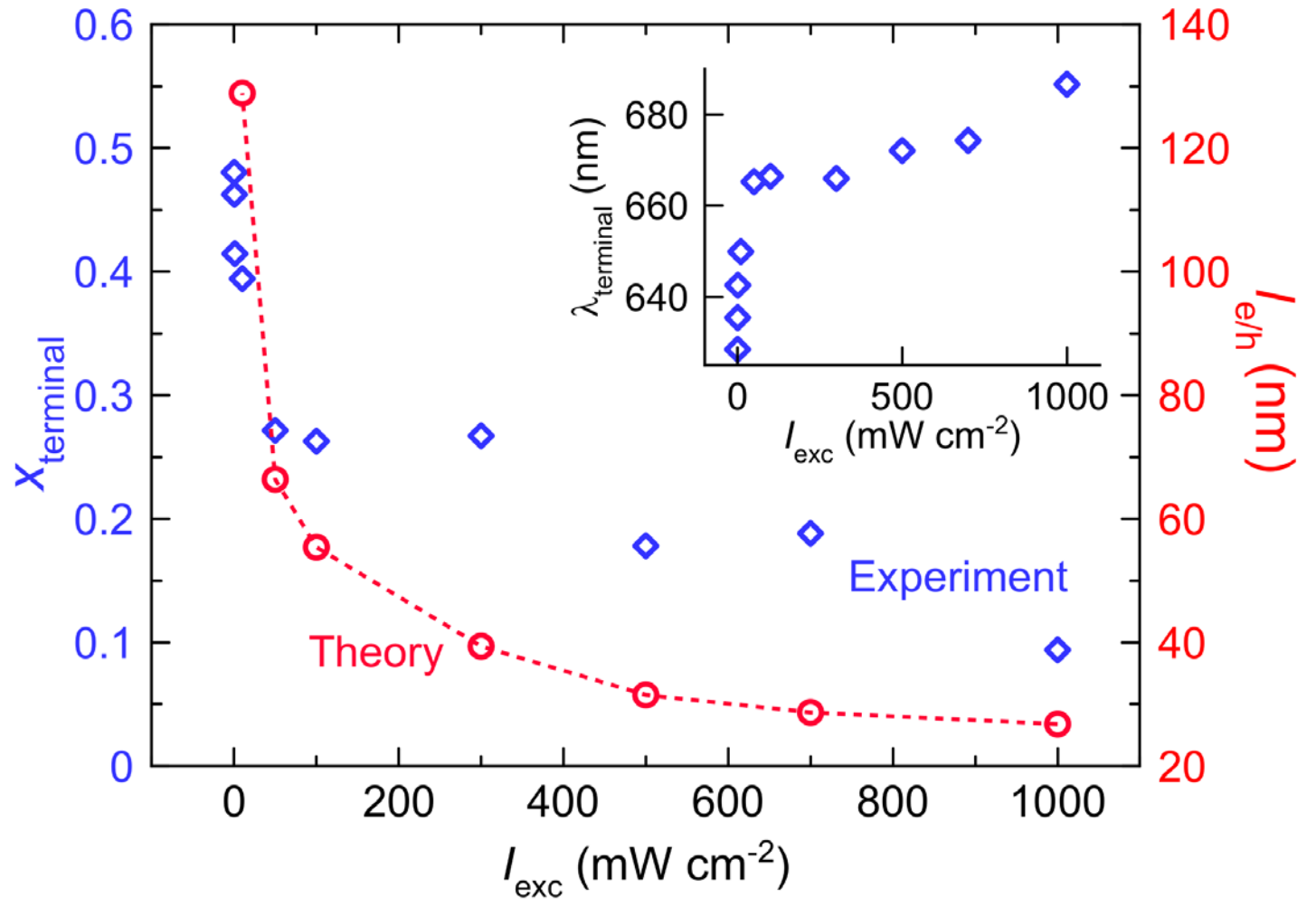

**Figure 5**. Bulk $CsPb(I_{0.5}Br_{0.5})_3$ thin film $x_{terminal}$-values (open blue diamonds) as functions of $I_{exc}$. Inset: Terminal peak wavelength ($\lambda_{terminal}$) as a function of $I_{exc}$. Also shown are corresponding, theory-extracted carrier diffusion lengths (red circles with dashed red line as a guide to the eye). Data from Reference 36.

A key distinction between the band gap thermodynamic model and other models of photosegregation is its prediction of an excitation intensity threshold ($I_{exc,threshold}$) below which no photosegregation occurs. Existence of $I_{exc,threshold}$ stems from the need to have a sufficient carrier number, $n$, to drive $\Delta F^*$ towards demixing. By rearranging **Equation 19**, $I_{exc,threshold}$ can be defined as

$$I_{\mathrm{exc,threshold}} = \frac{\Delta x}{x_{\mathrm{init}}(1-x_{\mathrm{init}})}\left[\frac{1}{\tau}\left(\frac{h\nu}{\alpha V_{\mathrm{D}}}\right)\left(\frac{kT-2U_{\mathrm{I,Br}}x_{\mathrm{init}}(1-x_{\mathrm{init}})}{\Delta E_{\mathrm{g,domain}}}\right)\right] \quad (20)$$

with $\Delta x = (x_{\mathrm{init}} - x_{\mathrm{terminal}})$ a minimum (practical) observable change in $x$ due to photosegregation.

**Figure 6** shows experimental $I_{exc,threshold}$-values (blue diamonds) for $CsPb(I_{1-x}Br_x)_3$ nanocrystals. The data reveal $I_{exc,threshold}$ to decrease with increasing $l_{nanocrystal}$. This is expected as $V_D = l_{\mathrm{e/h}}^3$ with $l_{nanocrystal}=l_{e/h}$. Plotted next to the experimental data are theoretical predictions for $\Delta x=0.01x_{init}$ using $\tau$-values, first employed in **Figure 4**. Accompanying theoretical $I_{exc,threshold}$-values (red circles) decrease with increasing $I_{exc}$ but again do not quantitatively agree with experiment. This stems from abovementioned limitations of the model, which lead to discrepancies between experiment and theory.

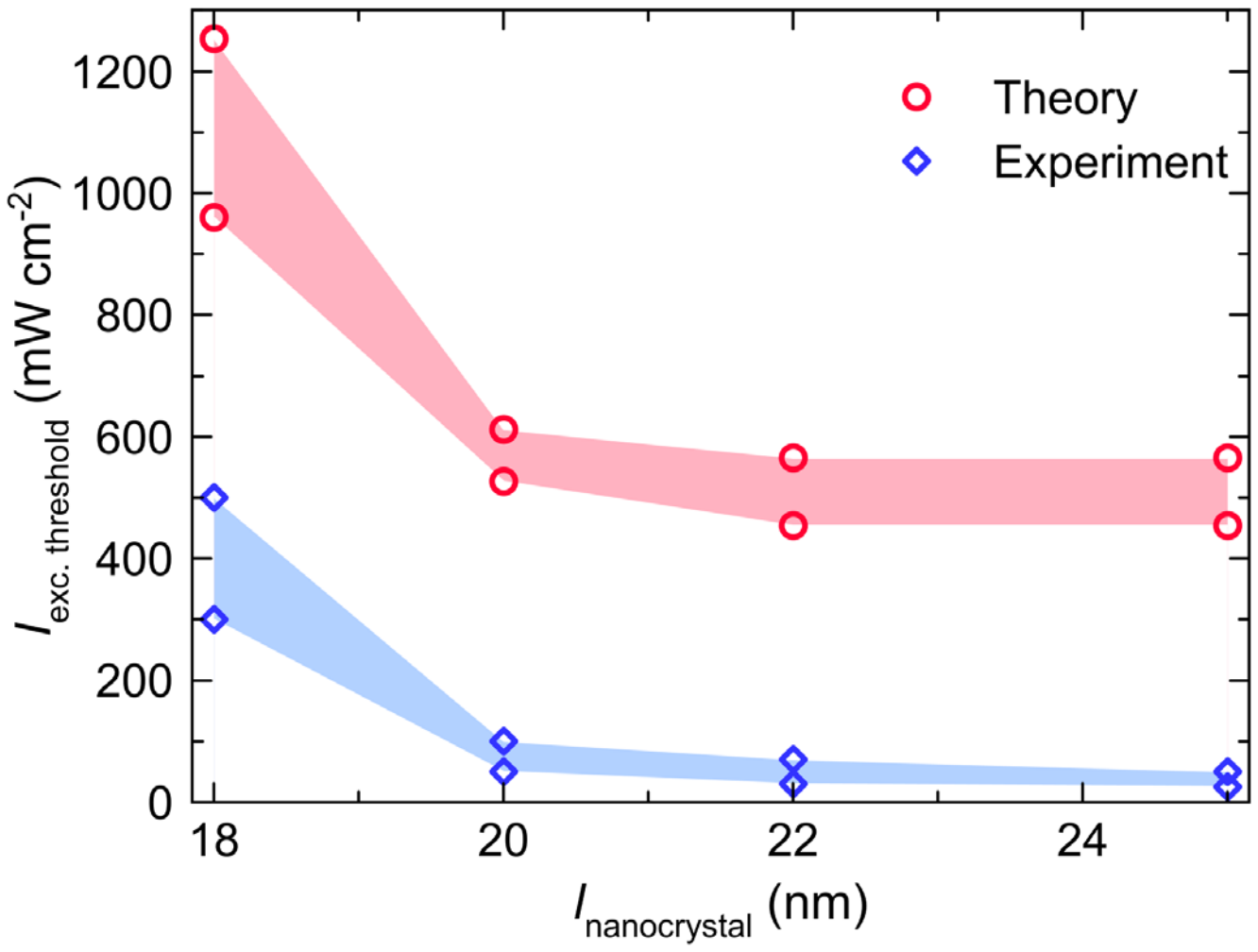

**Figure 6**. Experimental $I_{exc,threshold}$-ranges for $l_{nanocrystal}$=18, 20, 22, and 25 nm $CsPb(I_{0.5}Br_{0.5})_3$ nanocrystals (blue diamonds and shaded region). Theoretical $I_{exc,threshold}$ $I_{exc}$-dependency from a band gap-based thermodynamic model for photosegregation (red circles and shaded region). Data from Reference 36.

Predicting $I_{exc,threshold}$ for bulk materials is complicated by ambiguities in $l_{e/h}$. However, assuming nominal $l_{e/h}$-values between 100-1000 nm[19,36,40,42], order of magnitude $I_{exc,threshold}$-values range from 20 to 200 μW cm$^{-2}$.[19,20] These $I_{exc,threshold}$-values agree well with reported bulk $I_{exc,threshold}$-values for $MAPb(I_{1-x}Br_x)_3$ [$CsPb(I_{1-x}Br_x)_3$] of 30-500 μW cm$^{-2}$ [200 μW cm$^{-2}$].[16,20,35,36] Bulk values are smaller than those for nanocrystals, as expected. **Table 3** summarizes literature-reported $I_{exc,threshold}$-values.

**Table 3**. $I_{exc,threshold}$-values for mixed-halide perovskites.

| System | $I_{exc,threshold}$ (mW cm$^{-2}$) | Reference |
|---|---|---|
| $MAPb(Br_{0.5}I_{0.5})_3$ thin film | 0.04 | 19 |
| | 0.03-0.1 | 35 |
| $MAPb(Br_{0.6}I_{0.4})_3$ thin film | 0.1-0.23 | 20 |
| | 0.5 | 58 |
| $CsPb(Br_{0.5}I_{0.5})_3$ nanocrystal film | 1-500 | 36 |
| $CsPb(Br_{0.5}I_{0.5})_3$ thin film | 0.2 | 36 |
| $FA_{0.83}Cs_{0.17}(Br_{0.6}I_{0.4})_3$ thin film | 3 | 59 |
| $FA_{0.83}Cs_{0.17}(Br_{0.2}I_{0.8})_3$ thin film | 50 | 60 |

**Kinetics**

To complement the above thermodynamic formulation of the band gap model, photosegregation dynamics have been modeled using kinetic Monte Carlo (KMC) simulations.[61] KMC is a Markov-chain Monte Carlo method that excels at sampling discrete, dynamic processes. For the problem at hand, halide sites on a perovskite lattice represent a discrete network available for halide migration via hopping. KMC simulations therefore calculate reaction rates for all conceivable hops, taking into account the propensity for light-driven photosegregation in tandem with entropic influences due to local concentrations of reactive species. What results is an effective reaction path that incorporates the impact each of the band gap model's parameters has in determining the evolution of the parent, mixed-halide material under illumination. For researchers interested in running these simulations themselves, **SI Note 4** provides an example simulation.

Microscopically, halide exchange between an I-rich inclusion and its parent, mixed-halide phase involves a mediator. This is assumed to be a halide vacancy, $V_X$.[20] Vacancy involvement is supported by perovskite defect formation energies that readily produce Schottky defect pairs.[62] The hypothesis has been tested by monitoring photosegregation kinetics in halide-deficient, mixed-halide films with known vacancy concentrations ($y$).[20]

**Figure 7a** illustrates the $y$-dependency of predicted KMC photosegregation rate constants. **Figure 7b** shows accompanying, experimental (emission-based) rate constants. Linear $y$ dependencies in both cases strongly point to vacancy-mediated anion migration. Of note is that excess halide precursor feed ratios lower forward photosegregation rate constants. This suggests vacancy-mediated transport even at superstoichiometric halide concentrations.[20]

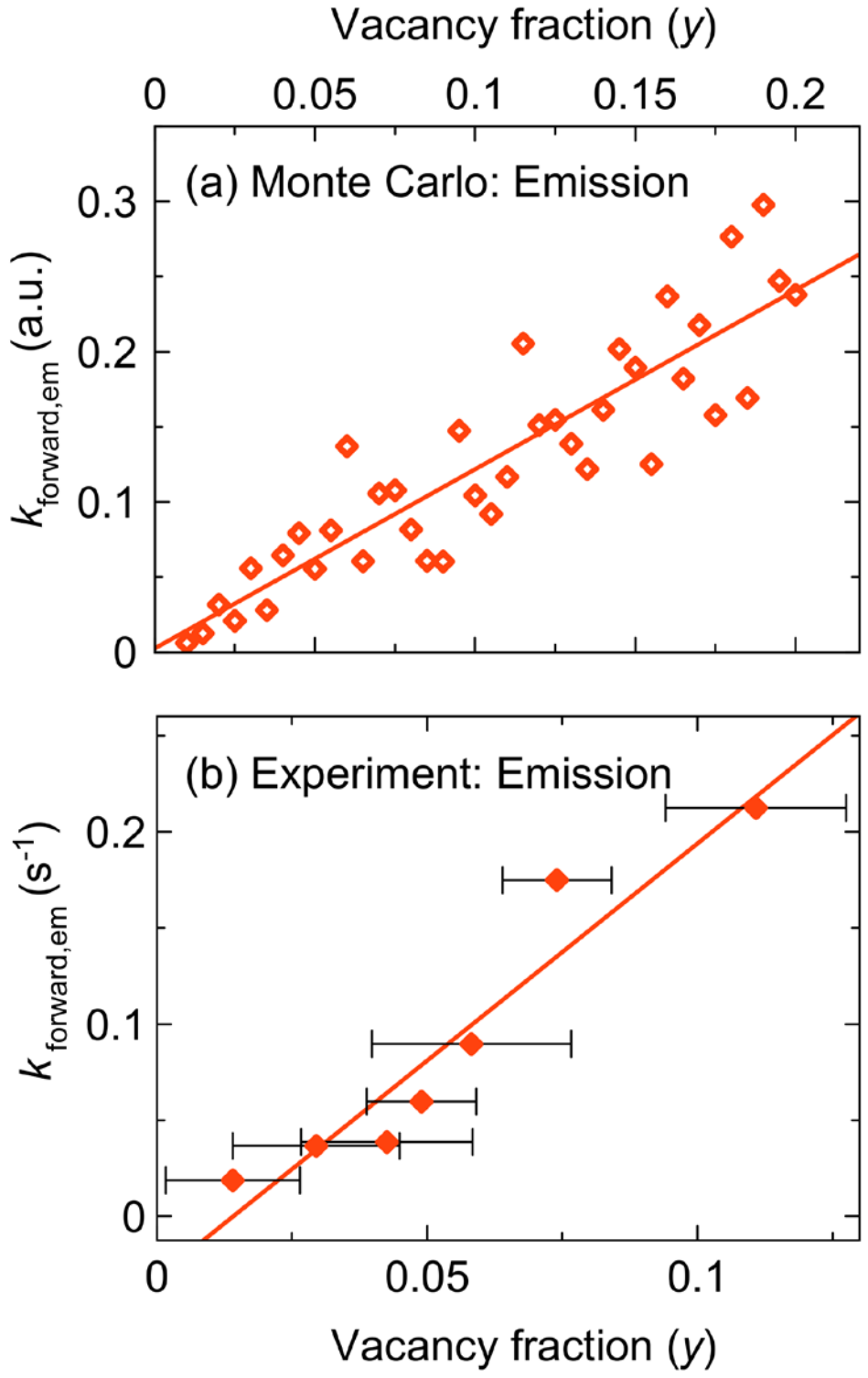

**Figure 7**. Dependency of emission-based, first order photosegregation rate constants with halide vacancy concentration, *y*, in (a) KMC and (b) experimental studies. Data from Reference 20.

**Figure 7** suggests that **Equation 13** can be expressed in terms of the following partial reactions that depend heavily on $V_X$ concentration

$$m\ \mathrm{APb}\left(\mathrm{I}_{1-x_{\mathrm{init}}}\mathrm{Br}_{x_{\mathrm{init}}}\right)_3 + V_X \leftrightarrow m\ \mathrm{APb}\left(\mathrm{I}_{1-x_{\mathrm{init}-\delta x'}}\mathrm{Br}_{x_{\mathrm{init}}}\right)_3 + \mathrm{I}^- \quad (21a)$$

$$m\ \mathrm{APb}\left(\mathrm{I}_{1-x_{\mathrm{init}-\delta x'}}\mathrm{Br}_{x_{\mathrm{init}}}\right)_3 + \mathrm{Br}^- \leftrightarrow m\ \mathrm{APb}\left(\mathrm{I}_{1-x_{\mathrm{init}-\delta x'}}\mathrm{Br}_{x_{\mathrm{init}+\delta x'}}\right)_3 + V_X \quad (21b)$$

$$j\ \mathrm{APb}(\mathrm{I}_{1-x}\mathrm{Br}_x)_3 + V_X \leftrightarrow j\ \mathrm{APb}(\mathrm{I}_{1-x}\mathrm{Br}_{x-\delta x})_3 + \mathrm{Br}^- \quad (21c)$$

$$j\ \mathrm{APb}(\mathrm{I}_{1-x}\mathrm{Br}_{x-\delta x})_3 + \mathrm{I}^- \leftrightarrow j\ \mathrm{APb}(\mathrm{I}_{1-x+\delta x}\mathrm{Br}_{x-\delta x})_3 + V_X. \quad (21d)$$

For simplicity, halide charges as well as those of their vacancies have been neglected. It is evident that **Equations 21 a** and **c** are rate-limiting due to their linear $V_X$ dependencies. This rationalizes both KMC and experimental behavior seen in **Figure 7**.

In practice, kinetic manifestations of **Equations 21 a-d**, are captured by emission-based photosegregation rate constants of the form[33,35]

$$k_{\mathrm{forward,em}}(I_{\mathrm{exc}}) = Ae^{-\frac{\left[-\left|\Delta E_{\mathrm{light}}(I_{\mathrm{exc}})\right| + E_{\mathrm{a,activation}}\right]}{RT}} \quad (22)$$

where $A$ is an attempt prefactor and $E_{\mathrm{a,activation}}$ is an anion hopping barrier of order 0.55 eV (54 kJ mol$^{-1}$).[32,63,64,65,66,67] $\Delta E_{\mathrm{light}}$ is an empirical parameter, linked to energetic contributions from photogenerated carriers that undergo localization in I-rich inclusions of the alloy. Under illumination, $\Delta E_{\text{light}}$ suppresses $E_{\text{a,activation}}$ to enhance anion migration. **Figures 8a** and **8b**, illustrate experimental and KMC $k_{\text{forward,em}}$-values, highlighting their $I_{\text{exc}}$-dependencies.

A corresponding entropically-driven (dark) remixing rate constant is

$$k_{\text{reverse}} = A' e^{-\frac{E_{\text{a,activation}}}{RT}} \tag{23}$$

where $A'$ is a dark attempt frequency. **Equations 22** and **23** collectively rationalize asymmetries in experimental photosegregation and entropically-driven remixing timescales. Specifically, when $A \approx A'$, $k_{\text{forward,PL}} > k_{\text{reverse}}$, which accounts for experimental photosegregation (dark remixing) timescales of order minutes (tens of minutes).[10,35,68]

**Equations 22** and **23** further rationalize photosegregation/dark remixing temperature dependencies.[10,16,35,30,69,70,72] For $MAPb(I_{0.5}Br_{0.5})_3$, **Figure 8c** summarizes activation energies, obtained using an Arrhenius analysis for temperatures between 320 K and 185 K. Extracted $E_{\text{a,activation}}$-values for $x_{\text{init}}$=0.4 and 0.5 are $E_{\text{a,activation}}$=0.27 eV (26 kJ mol$^{-1}$) and 0.30 eV (29 kJ mol$^{-1}$).[16] An analogous analysis of $x_{\text{init}}$=0.5 dark remixing data for temperatures between 295 K and 353 K (**Figure 8d**) yields $E_{\text{a,activation}}$=0.55 eV (54 kJ mol$^{-1}$).[35]

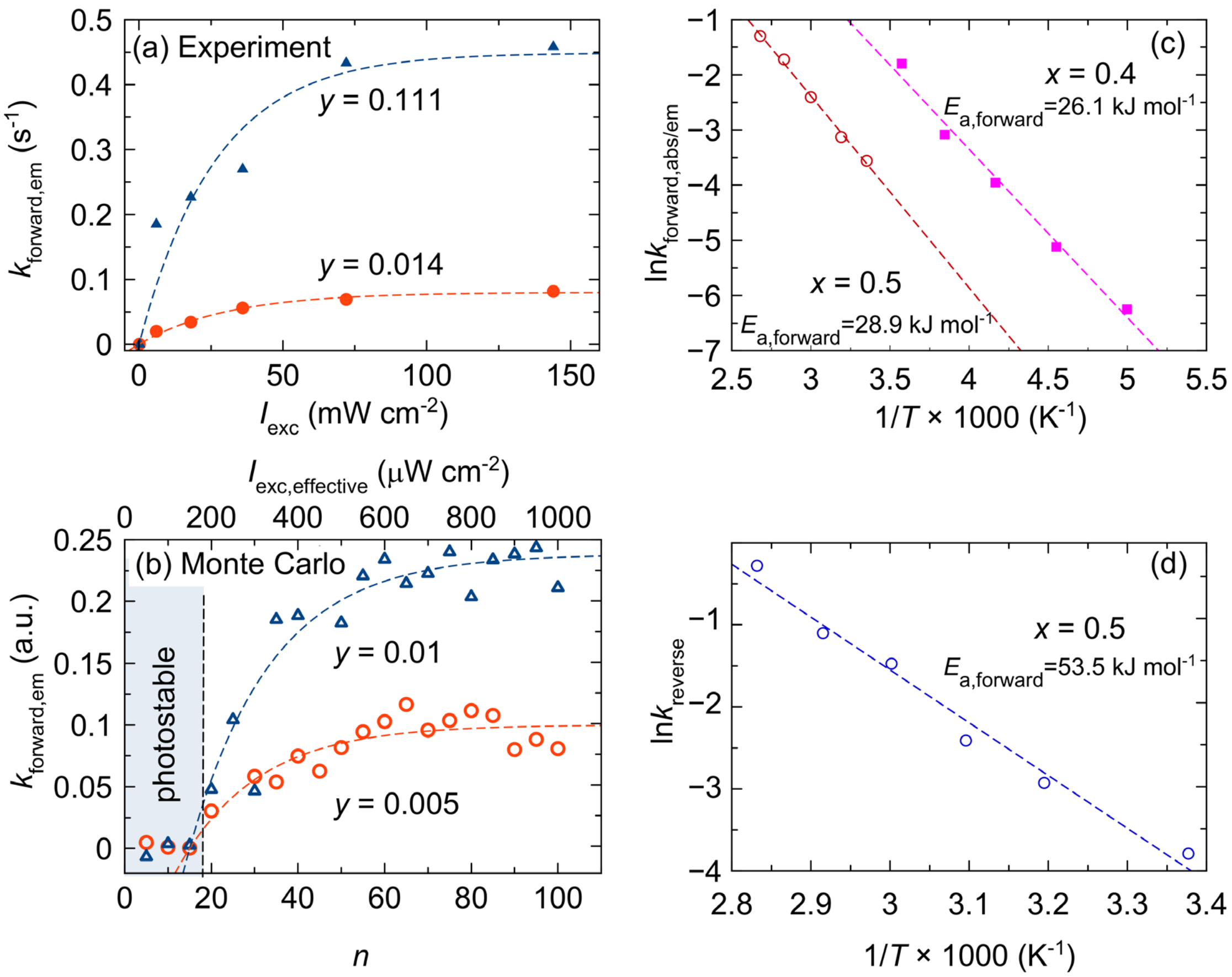

**Figure 8**. (a) Experimental and (b) KMC forward photosegregation rate constants versus $I_{\text{exc}}$ for two different vacancy concentrations. Dashed lines are rising exponential fits to $k_{\text{foward,em}}$. (c, d) Arrhenius-like temperature dependence of (c) $k_{\text{forward}}$ and (d) $k_{\text{reverse}}$. Dashed lines are linear fits to the data. Data from References 20 and 33.

**Equations 21 a-d** point to photosegregation being a stepwise process, potentially controlled via a single step. This is corroborated by high $I_{exc}$ KMC simulations using Maxwell-Boltzmann carrier statistics, which show I-rich domain growth via sequential $I^-$ and $Br^-$ exchange with surrounding, mixed-halide phases. In simulations, all photogenerated carriers localize in stochastically-generated I-rich domains, having a singular emission energy. Domains remain stationary in the lattice while their energies progressively decrease due to $I^-/Br^-$ exchange.

**Figure 9a** illustrates a resulting I-rich domain. **Figure 9b** is a corresponding radial distribution function that highlights $Br^-$ expulsion from the core and its subsequent accumulation at the inclusion's periphery. Some $Br^-$ remains trapped in the domain core due to a depletion of local vacancies that mediate $I^-/Br^-$ exchange. This results in finite $x_{terminal}$-values of order 0.2. Observed KMC $x_{terminal}$-values have a purely kinetic, as opposed to thermodynamic, origin. This is corroborated by high $I_{exc}$ (*i.e.*, large *n*) $MAPb(I_{1-x}Br_x)_3$ photosegregation measurements between 320 K and 185 K that show *T*-independent $x_{terminal}$-values (**Figure 9c**).[16]

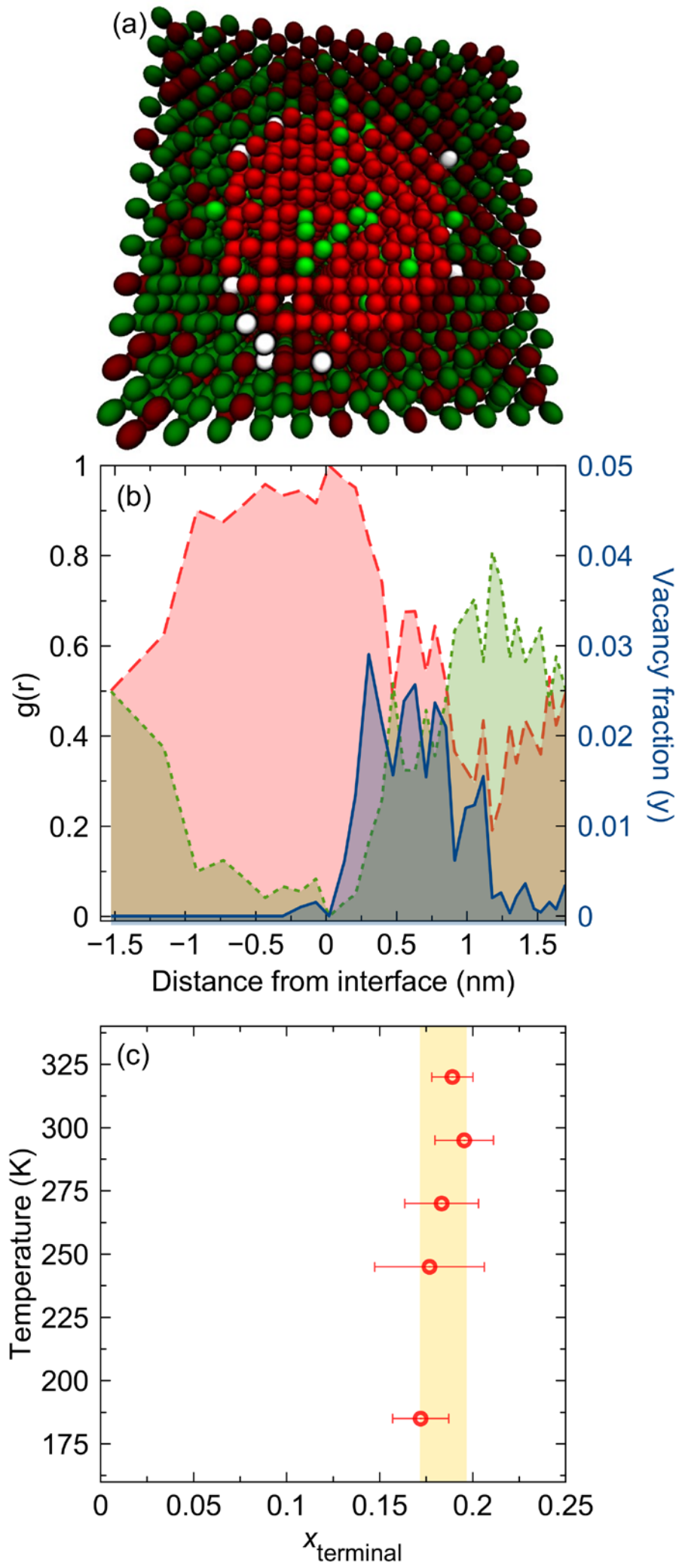

**Figure 9**. (a) I-rich domain resulting from high $I_{exc}$ Maxwell-Boltzmann-based KMC simulations of photosegregation in $MAPb(I_{1-x}Br_x)_3$. $I^-$ shown in red. $Br^-$ shown in green. $V_X$ shown in white. (b) Associated radial distribution functions for $I^-$ (red), $Br^-$ (green), and $V_X$ (blue) versus distance from the interface of the domain and the parent mixed-halide phase. (c) Experimental high $I_{exc}$ $MAPb(I_{1-x}Br_x)_3$ thin film $x_{terminal}$-values versus temperature for $x_{init}$=0.5. Data from References 16 and 20.

Implicit to measurements of $k_{forward,em}$ in **Figures 7** and **8** is the assumption that photosegregation reactions end with unique terminal stoichiometries (*e.g.*, **Equation 2**). In this limit, growth of redshifted emission follows exponential kinetics.[19,30,35,69] Recent compositionally-weighted X-ray diffraction measurements, however, now point to broad stoichiometric product distributions.[37] More importantly, they indicate that the photoluminescence growth kinetics being monitored are those of a ~1-2% minority fraction of I-enriched inclusions in the lattice. This suggests that the single exponential growth kinetics assumed until now are not formally correct.

To illustrate the complicated kinetics that underlie photosegregation, **Figure 10** summarizes large $I_{exc}$ KMC emission trajectories for a $x_{init}$=0.4 $MAPb(I_{1-x}Br_x)_3$ alloy undergoing photosegregation.[71] These KMC simulations include a more realistic accounting of photogenerated carrier occupation of I-rich inclusions through use of Fermi-Dirac statistics. Shown are four independent trajectories (dashed lines). Superimposed is the average trajectory (solid black line) that better depicts photosegregation's inherent kinetic heterogeneity.

**Figure 10** and like KMC/Fermi-Dirac simulations reveal three apparent stages of photosegregation. A first stage manifests itself as an induction time to observe I-rich domain growth. This has previously been reported by Herz *et al.* and is attributed to gradual increases in $I^-$ capture cross sections by growing I-rich domains.[72] A second stage reveals rapid compositional changes that lead to large spectral redshifts. This stage is likely what is observed and analyzed experimentally. The second stage behavior appears Avrami-like but also appears near exponential. A third growth stage leads to saturation of compositional changes and rationalizes terminal emission wavelength/energy saturation behavior seen in experiments and in **Figure 3**.

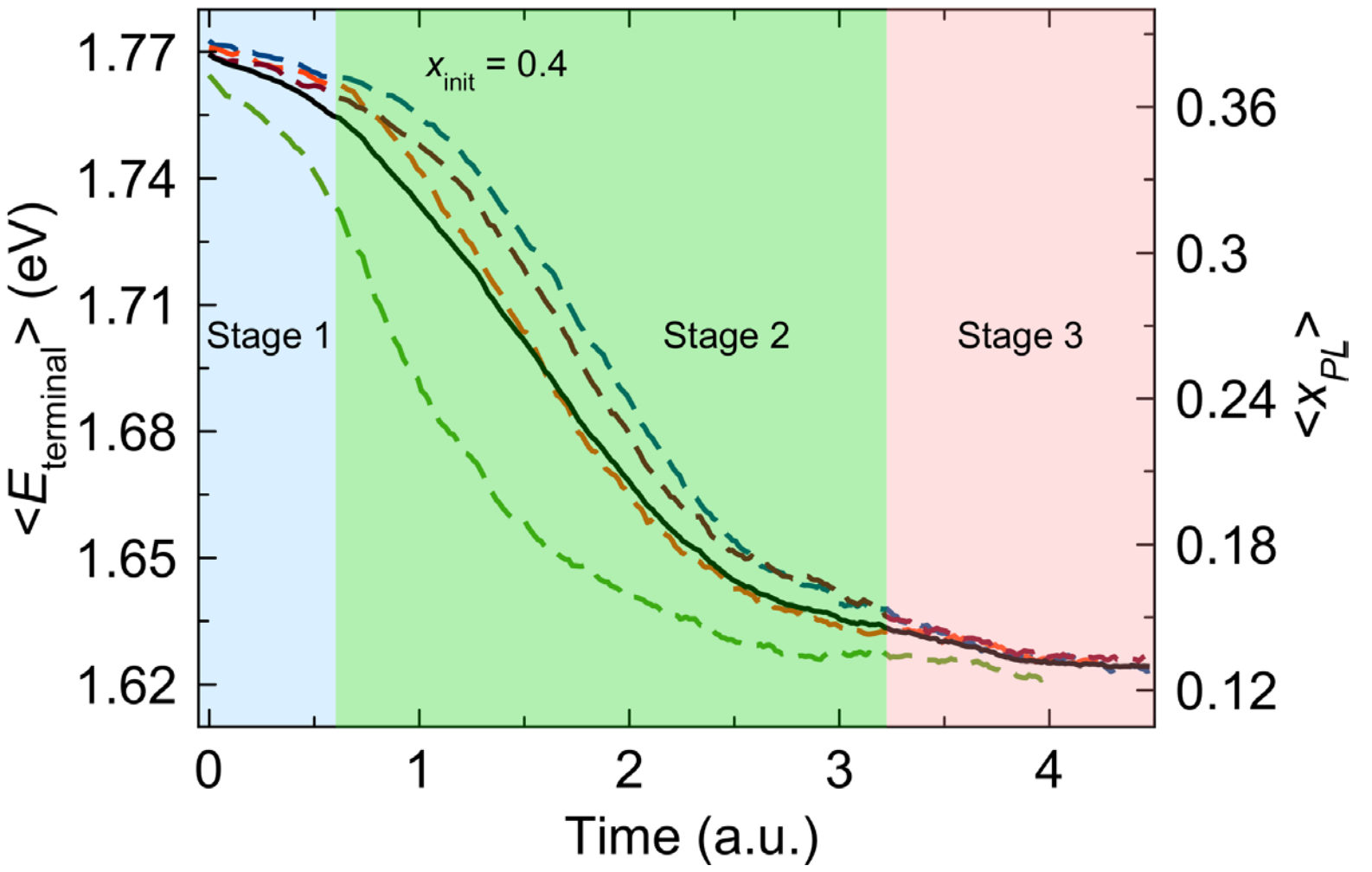

**Figure 10**. Four KMC/Fermi-Dirac trajectories (dashed lines) that plot the average emission-based stoichiometry, $<x_{PL}>$, present during simulations for $x_{init}$=0.4. The elapsed time is a combination of the number of steps taken and energy changes involved during those steps. $<x_{PL}>$ has been extracted from average terminal photocarrier energies ($<E_{terminal}>$) using Vegard's law (**Table 1**). Solid black line: Average behavior of the four trajectories. Data from Reference 71.

Of note is that these KMC/Fermi-Dirac simulations yield broad stoichiometric distributions that match those seen experimentally.[71] This highlights the importance of accounting for the photogenerated carrier occupation of I-rich domains, using Fermi-Dirac statistics. Moreover, a chemical potential for carriers in the system, $\varepsilon_{QFLS}$, emerges that quantifies their occupation of minority phases.

$\varepsilon_{\text{QFLS}}$ is consequential because it posits a new criterion for predicting $x_{\text{terminal}}$ and associated mixed-halide compositional photostabilities. Specifically, explicitly modeling the occupancy of product states in **Equation 2** with Fermi-Dirac statistics leads to the following alternative (low excitation intensity) expression for **Equation 17**[71]

$$\Delta F_{\text{light}} \cong \left(E_{\text{g},x} - E_{\text{g},x_{\text{init}}}\right) e^{-\left(\frac{E_{\text{g},x}-\varepsilon_{\text{QFLS}}}{kT}\right)}. \tag{24}$$

In **Equation 24**, $E_{\text{g},x}$ is the mixed-halide's composition-dependent band gap and $E_{\text{g},x_{\text{init}}}$ is the parent alloy's gap energy. **Equation 24** is derived in **SI Note 5**. What emerges is a free energy of mixing under illumination of

$$\Delta F^{*} \approx \Delta F_{\text{mix}} + \left(E_{\text{g},x} - E_{\text{g},x_{\text{init}}}\right) e^{-\left(\frac{E_{\text{g},x}-\varepsilon_{\text{QFLS}}}{kT}\right)}, \tag{25}$$

which is analogous to **Equation 18**. **Equation 25**, however, replaces $n$ with a statistical measure of I-rich domain occupation. The resulting $e^{-\left(\frac{E_{\text{g},x}-\varepsilon_{\text{QFLS}}}{kT}\right)}$ occupation factor leads to an enhanced sensitivity to local band gap energy differences, not captured by **Equation 18**. $\Delta F^{*}$ may therefore reconcile past differences between experiment and theory, especially regarding the band gap thermodynamic model's predictions of $x_{\text{terminal}}$ (**Figure 4**) and $I_{\text{exc,threshold}}$ (**Figure 6**).

Beyond this, $\varepsilon_{\text{QFLS}}$ and its accompanying $e^{-\left(\frac{E_{\text{g},x}-\varepsilon_{\text{QFLS}}}{kT}\right)}$ occupancy factor suggest a general rationalization for finite $x_{\text{terminal}}$ in $\text{MAPb(I}_{1-x}\text{Br}_x)_3$ and other mixed halide perovskites.[10,14,19,20,29,33,73,74,75] $\varepsilon_{\text{QFLS}}$ scales with the bandgap of the parent phase giving higher $x_{\text{init}}$ compositions a stronger propensity for carrier localization onto I-rich inclusions. Conversely, at low $x_{\text{init}}$ carrier trapping is insufficient to drive photosegregation. To illustrate, **Figure 11a** plots KMC-derived $\varepsilon_{\text{QFLS}}$-values across the $\text{MAPb(I}_{1-x}\text{Br}_x)_3$ compositional space.

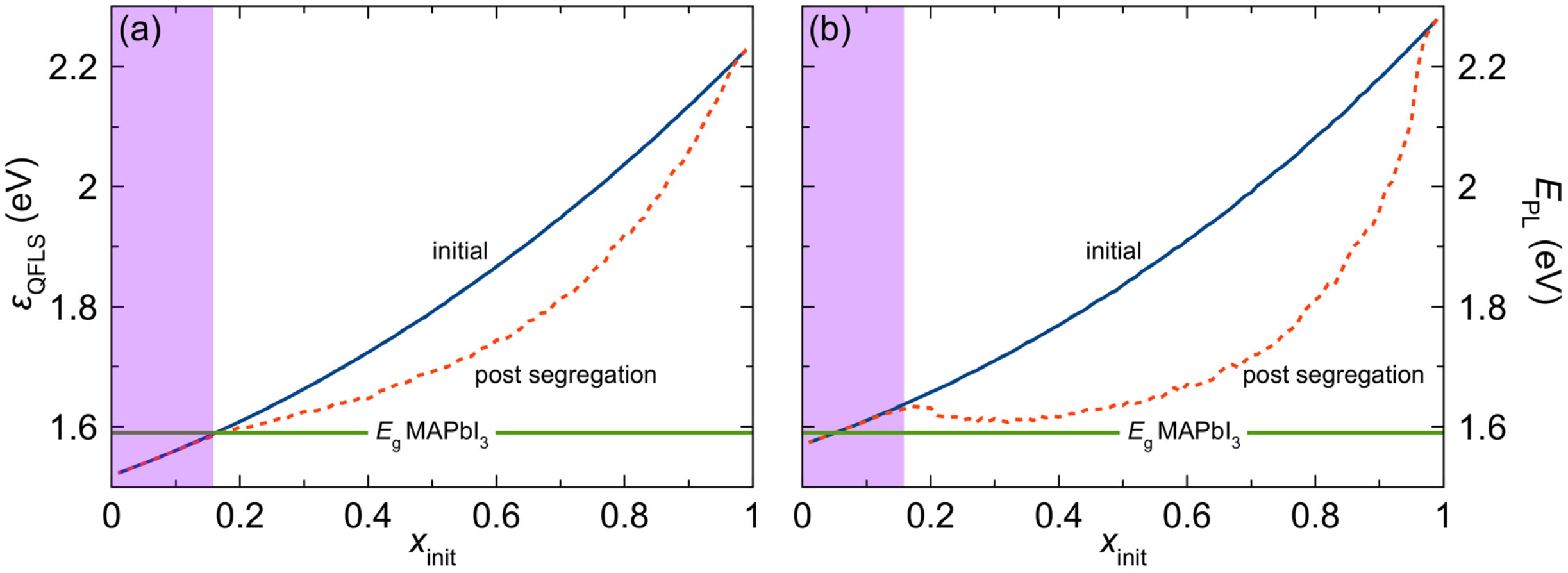

**Figure 11**. Plot of KMC-derived initial and post segregation (a) $\varepsilon_{\text{QFLS}}$ and (b) emission energy ($E_{\text{PL}}$) for different $x_{\text{init}}$ in $\text{MAPb(I}_{1-x}\text{Br}_x)_3$. Horizontal solid green lines denotes $E_g$ for $\text{MAPbI}_3$. Shaded violet region highlight compositions, $x \leq 0.2$, where little changes to $\varepsilon_{\text{QFLS}}$ or $E_{\text{PL}}$ are observed pre and post photosegregation. Data from Reference 71.

Evident is that $\varepsilon_{\text{QFLS}}$ decreases during photosegregation. This qualitatively reflects photogenerated carrier filling of I-rich domains accompanied by domain nucleation/growth.

**Figure 11b** shows accompanying changes to the emission energy ($E_{PL}$) which approaches that of pure $MAPbI_3$. $MAPbI_3$'s limiting $E_g$ is denoted with horizontal green lines in **Figure 11**.

Key in **Figure 11a** is the crossover between $\varepsilon_{QFLS}$ and $E_g$ $MAPbI_3$ when $x \leq 0.2$. The crossover implies a range of I-rich domain compositions where an insufficient driving force exists for their occupation by photogenerated carriers in the alloy. This is highlighted in **Figure 11b** by little to no change in $E_{PL}$ during photosegregation. Compositions below $\varepsilon_{QFLS}$~$E_g$ are therefore stable against photosegregation. This may explain why finite $x_{terminal}$~0.2 is commonly seen in $MAPb(I_{1-x}Br_x)_3$ and in other mixed-halide perovskites.

## Conclusion

In summary, a band gap thermodynamic model rationalizes most experimental photosegregation observations made in lead-based, mixed-halide perovskites. This includes: asymmetries in absorption and emission photosegregation response, $I_{exc}$- and $T$-dependent photosegregation rates and rate constants, $I_{exc}$-dependent and $T$-independent $x_{terminal}$, carrier diffusion length $x_{terminal}$ sensitivities, as well as the existence of $I_{exc,threshold}$. The above review quantitatively formulates the model to explicitly highlight its predictions. Recent amendments introduce the statistical occupation of photonucleated/photosegregated I-rich domains and may expand the model's predictive reach to reconcile remaining open questions on photochemical changes in mixed-halide perovskites as well as the highly intriguing observation of photoremixing at large excitation intensities.[76,77]

## Acknowledgements

We thank the Division of Materials Sciences and Engineering, Office of Basic Energy Sciences, U.S. Department of Energy (DOE) under Award DE-SC0014334 for financial support of this work. A.R. thanks the National Energy Research Scientific Computing Center (NERSC) for its generous support in the form of computing resources and guidance on high performance computing. We thank V. Trepalin for proofreading the manuscript.

Supplementary Information for:

**A thermodynamic band gap model for photoinduced phase segregation in mixed-halide perovskites**

Anthony Ruth[1], Halyna Okrepka[2], Prashant Kamat[2,3], Masaru Kuno[1,2]

1. University of Notre Dame, Department of Physics and Astronomy, Notre Dame, IN 46556, USA
2. University of Notre Dame, Department of Chemistry and Biochemistry and Department, Notre Dame, IN 46556, USA
3. Notre Dame Radiation Laboratory, Notre Dame, IN 46556, USA

## Supplementary Note 1. Chemical reactions of halide segregation and dark remixing

Halide photosegregation is first expressed as the photoinduced demixing of a mixed-halide alloy into anion-rich products. The alloy's initial stoichiometry is defined by $x_{\mathrm{init}}$, which for a mixed $I^-$/$Br^-$ alloy is its bromine fraction. What results is

$$\underset{\text{Alloy}}{\mathrm{APb}\left(\mathrm{I}_{1-\mathrm{x}_{\mathrm{init}}}\mathrm{Br}_{\mathrm{x}_{\mathrm{init}}}\right)_3} \leftrightarrow \underset{\text{I-rich}}{\frac{1}{2}\mathrm{APb}(\mathrm{I}_{1-\mathrm{x}}\mathrm{Br}_{\mathrm{x}})_3} + \underset{\text{Br-rich}}{\frac{1}{2}\mathrm{APb}\left(\mathrm{I}_{1-2\mathrm{x}_{\mathrm{init}}+x}\mathrm{Br}_{2\mathrm{x}_{\mathrm{init}}-\mathrm{x}}\right)_3} \tag{S1}$$

with A=methylammonium ($MA^+$) or formamidinium ($FA^+$) or $Cs^+$ and the first (second) product representing an I-rich (Br-rich) product phase. This is **Equation 1** of the main text. Equal quantities of I-enriched and Br-enriched inclusions are assumed to simplify the analysis. Further simplification comes from assuming that the parent alloy forms from disequal amounts of pure iodide and pure bromide species, *i.e.*,

$$\underset{\text{Pure I-phase}}{(1-x_{\mathrm{init}})\mathrm{APbI}_3} + \underset{\text{Pure Br-phase}}{x_{\mathrm{init}}\mathrm{APbBr}_3} \rightarrow \underset{\text{Alloy}}{\mathrm{APb}\left(\mathrm{I}_{1-\mathrm{x}_{\mathrm{init}}}\mathrm{Br}_{\mathrm{x}_{\mathrm{init}}}\right)_3}. \tag{S2}$$

Together, **Equations S1** and **S2** yield the net demixing reaction

$$\underset{\text{Pure I-phase}}{(1-x_{\mathrm{init}})\mathrm{APbI}_3} + \underset{\text{Pure Br-phase}}{x_{\mathrm{init}}\mathrm{APbBr}_3} \leftrightarrow \underset{\text{I-rich}}{\frac{1}{2}\mathrm{APb}(\mathrm{I}_{1-x}\mathrm{Br}_x)_3} + \underset{\text{Br-rich}}{\frac{1}{2}\mathrm{APb}\left(\mathrm{I}_{1-2x_{\mathrm{init}}+x}\mathrm{Br}_{2x_{\mathrm{init}}-x}\right)_3}, \tag{S3}$$

which is **Equation 2** of the main text. Derivations of mixing free energies for **Equation S3** are simplified because entropies and enthalpies of pure phases are zero.

**Alternative expressions**

**1A.** *Halide-symmetric reaction*

**Equation S3** simplifies further when using equal quantities of pure iodide and bromide phases (i.e., $x_{\mathrm{init}}$=0.5). This yields **Equation S4**

$$\tfrac{1}{2}\mathrm{APbI_3} + \tfrac{1}{2}\mathrm{APbBr_3} \leftrightarrow \tfrac{1}{2}\mathrm{APb(I_{1-x}Br_x)_3} + \tfrac{1}{2}\mathrm{APb(I_xBr_{1-x})_3} \quad \text{(S4)}$$

Pure I-phase  Pure Br-phase  I-rich  Br-rich whose symmetric products cancel antisymmetric energy terms in free energy calculations. **Equation S4** is therefore a useful starting point for understanding underlying thermodynamic forces, leading to photosegregation.

**1B.** *General reaction for all initial and final stoichiometries*

**Equation S3** does not permit certain combinations of $x_{\mathrm{init}}$ and $x$. For instance, the combination $x_{\mathrm{init}}$=0.8 and $x$=0.2 yields an unphysical stoichiometry for the $\mathrm{APb}\left(\mathrm{I}_{1-2x_{\mathrm{init}}+x}\mathrm{Br}_{2x_{\mathrm{init}}-x}\right)_3$ product. Namely, $2x_{\mathrm{init}} - x = 1.4$. **Equation S3** can therefore be made more general by assuming that its two products geometrically approach their pure phases via

$$(1 - x_{\mathrm{init}})\mathrm{APbI_3} + x_{\mathrm{init}}\mathrm{APbBr_3} \leftrightarrow (1 - x_{\mathrm{init}})\mathrm{APb(I_{1-x}Br_x)_3} + x_{\mathrm{init}}\mathrm{APb}\left(\mathrm{I}_{\frac{x}{x_{\mathrm{init}}}-x}\mathrm{Br}_{1+x-\frac{x}{x_{\mathrm{init}}}}\right)_3. \quad \text{(S5)}$$

Pure I-phase  Pure Br-phase  I-rich  Br-rich

Although more general, the $x_{\mathrm{init}}$ denominator in the $\mathrm{APb}\left(\mathrm{I}_{\frac{x}{x_{\mathrm{init}}}-x}\mathrm{Br}_{1+x-\frac{x}{x_{\mathrm{init}}}}\right)_3$ product leads to unwieldy expressions for corresponding free energies. For this reason, we prefer using **Equation S3**.

**1C.** *Reaction for minority phase formation*

An alternative approach to introducing variable product fractions into **Equation S3**, which avoids unphysical product stoichiometries, involves expressing its product stoichiometries in terms of $\xi$, the minority phase fraction. This assumes that the bromide-rich product, $\left[\tfrac{1}{2}\mathrm{APb}\left(\mathrm{I}_{1-2x_{\mathrm{init}}+x}\mathrm{Br}_{2x_{\mathrm{init}}-x}\right)_3\right]$, in **Equation S3** recombines with its parent phase to alter the parent phase stoichiometry. What results is

$$\xi\mathrm{APb(I_{1-x}Br_x)_3} + (1-\xi)\mathrm{APb}\left(\mathrm{I}_{1-x_{\mathrm{init}}}\mathrm{Br}_{x_{\mathrm{init}}}\right)_3 \leftrightarrow \mathrm{APb}\left(\mathrm{I}_{1-(x_{\mathrm{init}}+(x-x_{\mathrm{init}})\xi)}\mathrm{Br}_{x_{\mathrm{init}}+(x-x_{\mathrm{init}})\xi}\right)_3. \quad \text{(S6)}$$

Material of Br fraction $x$  Alloy  Modified alloy

**Equation S6** is expressed for the general case of a material of bromide fraction, $x$, that recombines with its parent phase.

**Equations S3** and **S6** can subsequently be combined to show the compositional change of the parent, mixed phase due to the formation of an iodine-rich phase with phase fraction, $\varphi$. Combining **Equations S3** and **S6** requires substituting $x$ with $2x_{\mathrm{init}}$-$x$ in **Equation S6**. What results is

$$\mathrm{APb}\left(\mathrm{I}_{1-x_{\mathrm{init}}}\mathrm{Br}_{x_{\mathrm{init}}}\right)_3 \leftrightarrow \varphi\mathrm{APb(I_{1-x}Br_x)_3} + (1-\varphi)\mathrm{APb}\left(\mathrm{I_{1-p}Br_p}\right)_3 \quad \text{(S7)}$$

Alloy  I-rich  Br-rich Alloy

where $p = \frac{x_{\text{init}} - \varphi x}{1-\varphi}$. **Equation S7** becomes convenient when examining the role of I-rich inclusion phase fractions in photosegregation.

**1D.** *Stepwise enrichment reaction*

A final approach to modeling photosegregation considers I-rich domains as discrete entities with finite sizes. In this treatment, $x$ and all properties that derive from it, depend locally on the domain's halide composition. The following equation shows the enrichment of a single domain by exchanging a single $Br^-$ with an $I^-$ from the parent phase

$$j\ \mathrm{APb(I_{1-x}Br_x)_3} + m\ \mathrm{APb}\left(\mathrm{I}_{1-x_{\text{init}}}\mathrm{Br}_{x_{\text{init}}}\right)_3 \leftrightarrow j\ \mathrm{APb(I_{1-x+\delta x}Br_{x-\delta x})_3} + m\ \mathrm{APb}\left(\mathrm{I}_{1-x_{\text{init}-\delta x'}}\mathrm{Br}_{x_{\text{init}+\delta x'}}\right)_3 \quad \text{(S8)}$$

The fractional enrichment of the domain is $\delta x = \frac{1}{3j}$ where $j = \frac{V_{\text{gap}}}{V_{\text{u.c.}}}$. The corresponding fractional enrichment of the parent phase is $\delta x' = \frac{1}{3m}$ where $m = \frac{V_D}{V_{\text{u.c.}}}$ and where $V_D = \frac{4}{3}\pi l_{\text{e/h}}^3$. The domain size can be pictured as a volume, $V_{\text{gap}}$ or as a length, $r_{\text{gap}} = (\frac{3}{4\pi} V_{\text{gap}})^{\frac{1}{3}}$, where local composition affects domain properties isotopically. Like **Equation S7**, **Equation S8** contains the minority phase fraction, $\varphi = \frac{j}{j+m}$. **Equation S8** is **Equation 13** of the main text.

By treating the $\mathrm{APb(I_{1-x+\delta x}Br_{x-\delta x})_3}$ I-rich phase in **Equation S8** as a discrete domain, its properties can be estimated via a cluster approximation based on the number of halide atoms in the cluster ($3j$). Cluster approximations have seen widespread use in condensed matter physics as ways to introduce and study local disorder in infinite lattices. Cluster approximations have been applied to electronic structure calculations[1,2] and to ionic interactions where they have significant use in the theory of binary alloys. The general quasichemical approximation (GQA)[3,4,5] allows one to estimate the phase fractions of multiple distinct phases, $x_i$, due to the free energy of each distinct composition. In the band gap thermodynamic model, a cluster approximation is therefore applied to the energy of band edge excitations, which are assumed to depend on a local band gap that is a function of $x_i$.

## Supplementary Note 2. Alternative expressions for ionic energy contributions to $\Delta F_{\text{mix}}$

Below are analytical expressions for free energy contributions to $\Delta F_{\text{mix}}$ in **Equations S1**, **S3**, **S4**, **S5**, **S7**, and **S8**. The change in Shannon mixing entropy, $-T\Delta S_{\text{mix}} = kT\sum_i x_i \ln(x_i)$ is evaluated for each equation. In all cases, $-T\Delta S_{\text{mix}}$ is Taylor series expanded to second order about $x=x_{\text{init}}$ to simplify resulting polynomial expressions.

Mixing enthalpy, $\Delta H_{\text{mix}} = U_{\text{I,Br}} \sum_i x_i (1 - x_i)$, contributions are also evaluated for each equation. $U_{\text{I,Br}}$ is an effective difference in attractive/repulsive strength between $I^-$ and $Br^-$ due to both Coulombic effects as well as lattice strain from halide mixing. Positive (negative) $U_{\text{I,Br}}$ values indicate net repulsion (attraction) between $I^-$ and $Br^-$.

**Table S1**. Compilation of energy contributions to $\Delta F_{mix}$.

| Equation | $-T\Delta S_{mix}$ | $-T\Delta S_{mix}$ (Approx.) | $\Delta H_{mix}$ |
|---|---|---|---|
| (S1) | $-3kT[x_{init}\ln(x_{init}) + (1-x_{init})\ln(1-x_{init})] + \frac{3}{2}kT[(2x_{init}-x)\ln(2x_{init}-x) + x\ln(x) + (1-x)\ln(1-x) + (1+x-2x_{init})\ln(1+x-2x_{init})]$ | $\frac{3kT}{2}\frac{(x-x_{init})^2}{x_{init}(1-x_{init})}$ | $-3U_{I,Br}(x-x_{init})^2$ |
| (S3) | $\frac{3}{2}kT[(2x_{init}-x)\ln(2x_{init}-x) + x\ln(x) + (1-x)\ln(1-x) + (1+x-2x_{init})\ln(1+x-2x_{init})]$ | $\frac{3kT}{2}\frac{(x-x_{init})^2 - x_{init}^2}{x_{init}(1-x_{init})}$ | $-3U_{I,Br}(x-x_{init})^2 + 3U_{I,Br}(x_{init}-x_{init}^2)$ |
| (S4) | $-3kT[x\ln(x) + (1-x)\ln(1-x)]$ | $6kT\{\left(x-\frac{1}{2}\right)^2 - \frac{1}{4}\}$ | $-3U_{I,Br}(x-0.5)^2 + \frac{3}{4}U_{I,Br}$ |
| (S5) | $-3kT\left[(1-x_{init})\{x\ln(x) + (1-x)\ln(1-x)\} + x_{init}\{\left(\frac{x}{x_{init}}-x\right)\ln\left(\frac{x}{x_{init}}-x\right) + \left(1+x-\frac{x}{x_{init}}\right)\ln\left(1+x-\frac{x}{x_{init}}\right)\}\right]$ | $\frac{3kT}{2}\frac{(x-x_{init})^2 - x_{init}^2}{x_{init}^2}$ | $3U_{I,Br}\frac{x}{x_{init}}(1-x_{init})(2x_{init}-x)$ |
| (S7) | $3kT\left[\varphi\{x\ln(x) + (1-x)\ln(1-x)\} + (1-\varphi)\left\{\frac{x_{init}-\varphi x}{1-\varphi}ln\left(\frac{x_{init}-\varphi x}{1-\varphi}\right) + \frac{1-\varphi-x_{init}+\varphi x}{1-\varphi}ln\left(\frac{1-\varphi-x_{init}+\varphi x}{1-\varphi}\right)\right\} - x_{init}\ln(x_{init}) - (1-x_{init})\ln(1-x_{init})\right]$ | $\frac{3kT}{2}\left(\frac{\varphi}{1-\varphi}\right)\frac{(x-x_{init})^2}{x_{init}(1-x_{init})}$ | $-3U_{I,Br}\frac{\varphi}{1-\varphi}(x-x_{init})^2$ |
| (S8) | $3kT[\ln\left(\frac{x_{init}}{x}\right) + \ln\left(\frac{1-x}{1-x_{init}}\right)]$ | $3kT\frac{(x_{init}-x)}{x_{init}(1-x_{init})}$ | $-6U_{I,Br}(x_{init}-x)$ |

## Supplementary Note 3. Derivation of Equation 12 in the main text.

This derivation originates from **SI** Reference 6. It assumes that compositional disorder in mixed-halide perovskites manifests itself as photoluminescence line broadening. Emission linewidths can therefore be used to estimate the degree of electronic disorder present in these materials.

The underlying degree of electronic disorder, $\delta E_g$, is modeled as

$$\delta E_{\mathrm{g}} = \delta x \left(\frac{dE_{\mathrm{g}}}{dx}\right) \quad \text{(S9)}$$

where $\delta x$ stands for variations in $x_i$ and $\left(\frac{dE_{\mathrm{g}}}{dx}\right)$, which is **Equation 16** of the main text, originates from a Vegard's law description of band gap compositional dependencies. $\delta x$ can be estimated by assuming that compositional disorder follows a Bernoulli process for the occupation of $j = \frac{V_{\mathrm{gap}}}{V_{\mathrm{u.c.}}}$ halide sites. This implies a binomial $x_i$ distribution. The probability of $Br^-$ occupying a halide site is $x_i$ and the probability of $I^-$ occupying a halide site is $1-x_i$. The associated standard deviation of the binomial distribution is then

$$\delta x = \frac{\sqrt{x(1-x)}}{\sqrt{j}}\,. \quad \text{(S10)}$$

Combining **Equations S9** and **S10** yields

$$V_{\mathrm{gap}} \geq \frac{V_{\mathrm{u.c.}}}{3}\left[\left(\frac{dE_{\mathrm{g}}}{dx}\right)\frac{\sqrt{x(1-x)}}{\delta E_{\mathrm{g}}}\right]^2, \quad \text{(S11)}$$

which is **Equation 12** of the main text.

## Supplementary Note 4. Examples of numerically simulating the band gap thermodynamic module using LattiKEM.

Numerical simulations of the bandgap thermodynamic model can be performed using the "mixedhalide" module of LattiKEM.[7] A basic settings file is provided below, which can be used as the basis to study the effects of $x_{\mathrm{init}}$, $I^-$ and $Br^-$ hopping barriers, temperature, carrier concentration, cluster size, stochastic effects, and $U_{I,Br}$.

```
sizex = 12                    #Supercell size
sizey = 12
sizez = 12
iodineRatio = 0.5             #This is 1-x
vacancyRatio = 0.005          #0.5 % will result in 25 vacancies
IBrRepulsiveEnergy = 0  #UI,Br
brHopEnergy = 0.25            #Eb,Br
iHopEnergy = 0.25             #Eb,I
temperature = 0.025           #kT in meV
numExcitations = 200          #Gives average carrier concentration of 0.115/formula unit
coordinationNumber = 5  #Cluster size
numSteps = 250000             #Number of KMC steps to perform
numRuns = 4                   #Perform 4 trajectories and average the results
averagingSteps = 1000   #Output will contain spectra for each 1000-step window
```

**Code block with example settings for LattiKEM to perform a photosegregation simulation**. 14 parameters are set, whilst the rest are determined by defaults. Text after the '#' are comments to explain the parameter in question.

Running the above simulation results in output containing the initial and final supercell, the total energy at each step of the simulation, and emission and absorption spectra for each 1000 step window (*i.e.*, steps 0-1000, steps 1000-2000, etc…).

The above simulation required approximately 6 days on a 256 core CPU. Runtime can be drastically reduced by taking advantage of the computational scaling which goes as O(Volume$^3$×Cluster size). Simulations can be performed in real time using supercells of around 6×6×6 formula units and a coordination number of 3. Further details can be seen in the example simulations included with the LattiKEM repository.

## Supplementary Note 5. Modeling Equation 24 of the main text.

We start by considering **Equation S1**,

$$\mathrm{APb}\left(\mathrm{I}_{1-x_{\mathrm{init}}}\mathrm{Br}_{x_{\mathrm{init}}}\right)_3 \leftrightarrow \frac{1}{2}\mathrm{APb}(\mathrm{I}_{1-x}\mathrm{Br}_x)_3 + \frac{1}{2}\mathrm{APb}\left(\mathrm{I}_{1-2x_{\mathrm{init}}+x}\mathrm{Br}_{2x_{\mathrm{init}}-x}\right)_3, \tag{S12}$$

which produces I-rich [$\mathrm{APb}(\mathrm{I}_{1-x}\mathrm{Br}_x)_3$] and Br-rich [$\mathrm{APb}\left(\mathrm{I}_{1-2x_{\mathrm{init}}+x}\mathrm{Br}_{2x_{\mathrm{init}}-x}\right)_3$] inclusions in a parent, mixed-halide film. $E_{\mathrm{g},x_{\mathrm{init}}}$, $E_{\mathrm{g},x}$, and $E_{\mathrm{g},2x_{\mathrm{init}}-x}$ are characteristic energy gaps of reactants and products.

The occupancy of product states is now evaluated using the Fermi-Dirac distribution, $w(E) = \frac{1}{1+e^{\frac{E-\varepsilon_{\mathrm{QFLS}}}{kT}}}$. This gives the following weighted sum for the electronic contribution to the free energy under illumination

$$\Delta F_{\mathrm{light}} = \left(E_{\mathrm{g},x} - E_{\mathrm{g},x_{\mathrm{init}}}\right)w\left(E_{\mathrm{g},x}\right) + \left(E_{\mathrm{g},2x_{\mathrm{init}}-x} - E_{\mathrm{g},x_{\mathrm{init}}}\right)w\left(E_{\mathrm{g},2x_{\mathrm{init}}-x}\right). \tag{S13}$$

The expression is simplified by asserting the low electron-hole occupancy of photoinduced Br-rich phases due to unfavorable band offsets. **Equation S13** simplifies to

$$\Delta F_{\text{light}} \approx \left(E_{g,x} - E_{g,x_{\text{init}}}\right)\left(\frac{1}{1+e^{\frac{E_{g,x}-\varepsilon_{\text{QFLS}}}{kT}}}\right). \tag{S14}$$

Since $E_{g,x} - \varepsilon_{\text{QFLS}} > kT$ at low excitation intensities, a further simplification, *i.e.*, $e^{\frac{E_{g,x}-\varepsilon_{\text{QFLS}}}{kT}} \gg 1$, can be made. This yields **Equation 24** of the main text.

The total free energy change of the system under illumination is therefore $\Delta F^* = \Delta F_{\text{mix}} + \Delta F_{\text{light}}$ or

$$\Delta F^* \approx \Delta F_{\text{mix}} + \left(E_{g,x} - E_{g,x_{\text{init}}}\right)e^{-\left(\frac{E_{g,x}-\varepsilon_{\text{QFLS}}}{kT}\right)}. \tag{S15}$$

This is **Equation 25** in the main text.

**SI References**